\documentclass[12pt]{article}

\usepackage{latexsym,amsmath,amscd,amssymb,graphics}
\usepackage{enumerate,framed}

\usepackage{graphicx}

\usepackage[square,authoryear]{natbib}
\usepackage{url}

\usepackage[all]{xy}

\makeatletter

\@addtoreset{figure}{section}
\def\thefigure{\thesection.\@arabic\c@figure}
\def\fps@figure{h, t}
\@addtoreset{table}{bsection}
\def\thetable{\thesection.\@arabic\c@table}
\def\fps@table{h, t}
\@addtoreset{equation}{section}

\makeatother

\begin{document}

\newtheorem{theorem}{Theorem}[section]
\newtheorem{definition}[theorem]{Definition}
\newtheorem{lemma}[theorem]{Lemma}
\newtheorem{remark}[theorem]{Remark}
\newtheorem{proposition}[theorem]{Proposition}
\newtheorem{corollary}[theorem]{Corollary}
\newtheorem{example}[theorem]{Example}

\def\below#1#2{\mathrel{\mathop{#1}\limits_{#2}}}



\title{Affine Lie-Poisson Reduction, Yang-Mills magnetohydrodynamics, and superfluids}
\author{Fran\c{c}ois Gay-Balmaz$^{1}$ and Tudor S. Ratiu$^{1}$}
\addtocounter{footnote}{1} \footnotetext{Section de
Math\'ematiques and Bernoulli Center, \'Ecole Polytechnique F\'ed\'erale de Lausanne.
CH--1015 Lausanne. Switzerland.
\texttt{Francois.Gay-Balmaz@epfl.ch, Tudor.Ratiu@epfl.ch}
\addtocounter{footnote}{1} }

\date{ }
\maketitle

\makeatother

\maketitle




\begin{abstract} This paper develops the theory of affine Lie-Poisson reduction and applies this process to Yang-Mills and Hall magnetohydrodynamics for fluids and superfluids. As a consequence of this approach, the associated Poisson brackets are obtained by reduction of a canonical cotangent bundle. A Kelvin-Noether circulation theorem is presented and is applied to these examples.
\end{abstract}



\section{Introduction}\label{Introduction}

The equations of motion of a non-relativistic \textit{adiabatic compressible fluid} are given by
\begin{equation}\label{ICAF}
\left\lbrace
\begin{array}{ll}
\vspace{0.2cm}\displaystyle\frac{\partial \mathbf{u}}{\partial
t}+\nabla_\mathbf{u}\mathbf{u}=\frac{1}{\rho}\operatorname{grad}p,\\
\vspace{0.2cm}\displaystyle\frac{\partial \rho}{\partial
t}+\operatorname{div}(\rho \mathbf{u})=0,\quad\frac{\partial S}{\partial t}+\operatorname{div}(S\mathbf{u})=0,
\end{array} \right.
\end{equation}
where $\rho$ is the mass density, $S$ is the entropy density, and $p$ is
the pressure. It was shown in \cite{MoGr1980} that this system, as well as its magnetohydrodynamic extension,
admit a noncanonical Poisson formulation, that is, equation \eqref{ICAF} can be written as
\[
\dot f=\{f,h\},
\]
relative to a Hamiltonian function $h$. The study of the relativistic case was initiated in \cite{BiHuTu1984}, \cite{Ma1984}, and \cite{HoKu1984a}.
The present paper considers only non-relativistic fluids.

It is of great (mathematical and physical) interest to obtain these Poisson brackets by a reduction procedure from a canonical Hamiltonian formulation on a cotangent bundle. In \cite{MaRaWe1984}, the noncanonical Poisson bracket associated to \eqref{ICAF} is obtained via Lie-Poisson
reduction for a semidirect product group involving the diffeomorphisms group of the fluid container and the space of the advected quantities $\rho$ and $S$. In the same spirit, the noncanonical Hamiltonian structure for \textit{adiabatic Yang-Mills charged fluid} discovered in \cite{GiHoKu1983} is obtained by reduction from a canonical formulation in \cite{GBRa2008a}, by using a Kaluza-Klein point of view involving the automorphism group of the principal bundle of the theory.

Noncanonical Hamiltonian structures for a wide class of non-dissipative fluid models were derived in \cite{HoKu1984b}, \cite{Ho1987}, \cite{HoKu1987}, \cite{HoKu1988} and \cite{Ho2001}. These examples include \textit{Yang-Mills magnetohydrodynamics}, \textit{spin glasses}, and various models of \textit{superfluids}, and involve Lie-Poisson brackets \textit{with cocycles}. Remarkably, from a mathematical point of view, the Hamiltonian structures of many of these models are identical. This Hamiltonian structure is studied in more detail, with an application to \textit{liquid crystals}, in \cite{Ho2002}. We will refer to all these models as \textit{complex fluids}.

In this paper we show the remarkable property that these Lie-Poisson brackets with cocycles can also be obtained by Poisson reduction from a canonical Hamiltonian structure.
The cocycle in the Hamiltonian structure appears only after reduction and it is due to the presence of an affine term added to the cotangent lifted action. The associated reduction process is naturally called \textit{affine Lie-Poisson reduction\/}.

An important example of such an affine action is given by the usual action of the automorphism group of a principal bundle on the connection forms. As a result we obtain, in a natural way, covariant differentials and covariant divergences in the expression of the Poisson brackets and of the reduced equations. These gauge theory aspects in the case of complex fluids are mathematically and physically interesting since they represent a bridge to other possible gauge theories in physics.

\medskip

We begin by recalling some needed facts about Lie-Poisson reduction for semidirect products (see \cite{HoMaRa1998}). Let $\rho : G\rightarrow \operatorname{Aut}(V)$ denote a \textit{right\/} Lie group representation of $G$ on the vector space $V$. As a set, the semidirect product $S=G\,\circledS\,V$ is the Cartesian product $S=G\times V$ whose group multiplication is given by
\[
(g_1,v_1)(g_2,v_2)=(g_1g_2,v_2+\rho_{g_2}(v_1)).
\]
The Lie algebra of $S$ is the semidirect product Lie algebra, $\mathfrak{s}=\mathfrak{g}\,\circledS\,V$, whose bracket has the expression
\[
\operatorname{ad}_{(\xi_1,v_1)}(\xi_2,v_2)=[(\xi_1,v_1),(\xi_2,v_2)]=([\xi_1,\xi_2],v_1\xi_2-v_2\xi_1),
\]
where $v\xi$ denotes the induced action of $\mathfrak{g}$ on $V$, that is,
\[
v\xi:=\left.\frac{d}{dt}\right|_{t=0}\rho_{\operatorname{exp}(t\xi)}(v)\in
V.
\]
From the expression for the Lie bracket, it follows that for $(\xi,v)\in\mathfrak{s}$ and $(\mu,a)\in\mathfrak{s}^*$ we have
\[
\operatorname{ad}^*_{(\xi,v)}(\mu,a)=(\operatorname{ad}^*_\xi\mu+v\diamond
a,a\xi),
\]
where $a\xi\in V^*$ and $v\diamond a\in\mathfrak{g}^*$ are given by
\[
a\xi:=\left.\frac{d}{dt}\right|_{t=0}\rho^*_{\operatorname{exp}(-t\xi)}(a)\quad\text{and}\quad
\langle v\diamond a,\xi\rangle_\mathfrak{g}:=-\langle a\xi,v\rangle_V,
\]
and where $\left\langle\cdot , \cdot \right\rangle_ \mathfrak{g}: \mathfrak{g} ^\ast \times \mathfrak{g}\rightarrow \mathbb{R}$ and $\left\langle \cdot , \cdot \right\rangle_V: V ^\ast \times V \rightarrow \mathbb{R}$ are the duality parings.
\medskip

\paragraph{Hamiltonian semidirect product theory.} Let $S := G
\,\circledS\,V $ be the semidirect product defined before. The lift of right translation of $S $ on $T ^\ast S $
induces a right action on $T ^\ast G \times V ^\ast$. Consider a Hamiltonian function $H:
T ^\ast G \times V ^\ast \rightarrow \mathbb{R}$ right
invariant under the $S $-action on $T ^\ast G \times V ^\ast$. In particular, the
function $H_{a_0}: = H|_{T ^\ast G\times \{a_0\}}: T ^\ast G
\rightarrow \mathbb{R}$ is invariant under the induced action
of the isotropy subgroup $G_{a_0}: = \{g \in G \mid \rho_g^\ast a_0 = a_0\}$
for
any $a_0 \in V ^\ast$. The following theorem is an easy consequence of the
semidirect product reduction theorem (see \cite{MaRaWe1984}) and the reduction
by stages method (see \cite{MaMiOrPeRa2007}).

\begin{theorem}\label{LPSD}
For $\alpha(t)\in T^*_{g(t)}G$ and
$\mu(t):=T^*R_{g(t)}(\alpha(t))\in\mathfrak{g}^*$, the following are
equivalent:
\begin{itemize}
\item[\bf{i}] $\alpha(t)$ satisfies Hamilton's equations for
$H_{a_0}$ on $T^*G$.
\item[\bf{ii}] The Lie-Poisson equation holds on $\mathfrak{s}^*$:
\[
\frac{\partial}{\partial t}(\mu,a)=-\operatorname{ad}^*_{\left(\frac{\delta
h}{\delta\mu},\frac{\delta h}{\delta a}\right)}(\mu,a)
=- \left(\operatorname{ad}^*_{\frac {\delta h}{ \delta \mu}}\mu+\frac {\delta
h}{ \delta a}\diamond
a,a\frac {\delta h}{ \delta \mu}\right),\quad a(0)=a_0
\]
where $\mathfrak{s}$ is the semidirect product Lie algebra
$\mathfrak{s}=\mathfrak{g}\,\circledS\, V$. The associated Poisson bracket is the Lie-Poisson bracket on the semidirect product Lie algebra $\mathfrak{s}^*$, that is,
\[
\{f,g\}(\mu,a)=\left\langle\mu,\left[\frac{\delta
f}{\delta\mu},\frac{\delta
g}{\delta\mu}\right]\right\rangle+\left\langle a,\frac{\delta f}{\delta
a}\frac{\delta g}{\delta\mu}-\frac{\delta g}{\delta a}\frac{\delta
f}{\delta\mu}\right\rangle.
\]
\end{itemize}
The evolution of the advected quantities is given by $a(t)=\rho^*_{g(t)}(a_0)$.
\end{theorem}


\section{Affine Lie-Poisson Reduction}\label{AMLPR}

The goal of this section is to carry out a generalization of the standard process of Lie-Poisson reduction for Lie groups, which is motivated by the example of superfluids. The only modification lies in the fact that the Lie group $G$ acts on its cotangent bundle by a cotangent lift \textit{plus an affine term}. The principal result of this section states that, under some conditions, reducing a canonical symplectic form relative to a cotangent lift with an affine term is equivalent to reduce a magnetic symplectic form relative to the right-cotangent lift. At the reduced level, we obtain affine Lie-Poisson brackets and affine coadjoint orbits, whose affine terms depend on the affine term in the action.

Consider the cotangent lift $R^{T^*}_g$ of the right translation $R_g$ on a Lie group $G$. Recall that $R^{T^*}_g$ is the right action of $G$ on $T^*G$ given by
\[
R^{T^*}_g(\alpha_f)=T^*R_{g^{-1}}(\alpha_f).
\]
Consider the map $\Psi_g :T^*G\rightarrow T^*G$ defined by
\begin{equation}\label{affine_cot_lift}
\Psi_g(\alpha_f):=R^{T^*}_g(\alpha_f)+C_g(f),
\end{equation}
where $C:G\times G\rightarrow T^*G$ is a smooth map such that $C_g(f)\in T^*_{fg}G$, for all $f,g\in G$. The map $\Psi_g$ is seen here as a modification of the cotangent lift by an affine term $C$. The following lemma gives the conditions guaranteeing that the map $\Psi_g$ is a right action.

\begin{lemma} Consider the map $\Psi_g$ defined in \eqref{affine_cot_lift}. The following are equivalent.
\begin{itemize}
\item[\bf{i}] $\Psi_g$ is a right action.
\item[\bf{ii}] For all $f,g,h\in G$, the affine term $C$ verifies the property
\begin{equation}\label{cocycle_property}
C_{gh}(f)=C_h(fg)+R^{T^*}_h(C_g(f)).
\end{equation}
\item[\bf{iii}] There exists a one-form $\alpha\in\Omega^1(G)$ such that $C_g(f)=\alpha(fg)-R^{T^*}_g(\alpha(f))$.
\end{itemize}
\end{lemma}

We denote by $\mathcal{C}(G)$ the space of all maps $C : G\times G\rightarrow T^*G,\;(g,f)\mapsto C_g(f)\in T^*_{fg}G$ verifying the property \eqref{cocycle_property}.
Remark that given an affine term $C\in\mathcal{C}(G)$, the one-form $\alpha$ in item $\bf{iii}$, is only determined up to a right-invariant one-form. Denoting by $\Omega^1_R(G)$ the space of all right-invariant one-forms on $G$, we have an isomorphism between $\mathcal{C}(G)$ and $\Omega^1(G)/\Omega^1_R(G)$. This space is clearly isomorphic to the space $\Omega^1_0(G)$ of all one-forms $\alpha$ on $G$ such that $\alpha(e)=0$. We can now state the main result of this section.

\medskip

\begin{theorem}\label{affine_LiePoisson_reduction} 
Consider the symplectic manifold $(T^*G,\Omega_{\rm can})$, and the affine action
\[
\Psi_g(\beta_f):=R^{T^*}_g(\beta_f)+C_g(f),
\]
where $C\in\mathcal{C}(G)$. Let $\alpha\in\Omega^1_0(G)$ be the one-form associated to $\Psi_g$. Then the following hold:
\begin{itemize}
\item[\bf{i}] The fiber translation $t_\alpha : (T^*G,\Omega_{\rm can})\rightarrow (T^*G,\Omega_{\rm can}-\pi^*_G\mathbf{d}\alpha)$ is a symplectic map. The action induced by $\Psi_g$ on $(T^*G,\Omega_{\rm can}-\pi^*_G\mathbf{d}\alpha)$ through $t_\alpha$ is simply the cotangent lift $R^{T^*}_g$.
\item[\bf{ii}] Suppose that $\mathbf{d}\alpha$ is $G$-invariant. Then the action $\Psi_g$ is symplectic relative to the canonical symplectic form $\Omega_{\rm can}$.
\item[\bf{iii}] Suppose that there is a smooth map $\phi : G\rightarrow \mathfrak{g}^*$ that satisfies
\[
\mathbf{i}_{\xi^L}\mathbf{d}\alpha=\mathbf{d}\langle\phi,\xi\rangle
\]
for all $\xi\in\mathfrak{g}$, where $\xi^L$ is the left invariant extension of $\xi$. Then the map
\[
\mathbf{J}_\alpha=\mathbf{J}_R\circ t_\alpha-\phi\circ\pi_G,
\]
where $\mathbf{J}_R(\alpha_f)=T^*_eL_f(\alpha_f)$, is a momentum map for the action $\Psi_g$ relative to the canonical symplectic form. We can always choose $\phi$ such that $\phi(e)=0$. In this case, the nonequivariance cocycle of $\mathbf{J}_\alpha$ is $\sigma=-\phi$.
\item[\bf{iv}] The symplectic reduced space $(\mathbf{J}_\alpha^{-1}(\mu)/G_\mu^{\,\sigma},\Omega_\mu)$ is symplectically diffeomorphic to the affine coadjoint orbit
\[
\mathcal{O}^{\,\sigma}_\mu=\left\{\operatorname{Ad}^*_g\mu+\sigma(g)\mid g\in G\right\},
\]
endowed with the affine orbit symplectic form
\begin{align*}
\omega^+_\sigma(\lambda)\left(\operatorname{ad}^*_\xi\lambda\right.&-\Sigma(\xi,\cdot),\left.\operatorname{ad}^*_\eta\lambda-\Sigma(\eta,\cdot) \right)\\
&=\langle\lambda,[\xi,\eta]\rangle-\Sigma(\xi,\eta),
\end{align*}
where $\Sigma(\xi,\cdot):=-T_e\sigma(\xi)$. The symplectic diffeomorphism is induced by the $G_\mu^{\,\sigma}$-invariant smooth map
\[
\psi : \mathbf{J}_\alpha^{-1}(\mu)\rightarrow\mathcal{O}^{\,\sigma}_\mu,\quad\psi(\alpha_g):=\Psi_{g^{-1}}(\alpha_g).
\]
\end{itemize}
\end{theorem}

The affine coadjoint orbits $(\mathcal{O}^{\,\sigma}_\mu,\omega^+_\sigma)$ are symplectic leaves in the affine Lie-Poisson space $(\mathfrak{g}^*,\{\,,\}_\sigma^+)$, where
\begin{equation}\label{affine_Lie_Poisson}
\{f,g\}_\sigma^+(\mu)=\left\langle\mu,\left[\frac{\delta f}{\delta\mu},\frac{\delta g}{\delta\mu}\right]\right\rangle-\Sigma\left(\frac{\delta f}{\delta\mu},\frac{\delta g}{\delta\mu}\right).
\end{equation}


\section{Affine Hamiltonian Semidirect Product Theory}\label{AHSPT}

In this section we carry out the Poisson and symplectic reductions of a canonical cotangent bundle $(T^*S,\Omega_{\rm can})$, where $S=G\,\circledS\,V$ is the semidirect product of a Lie group $G$ and a vector space $V$ and where $S$ acts on its cotangent bundle by cotangent lift \textit{plus an affine term}. We will see that this process is a particular case of the theory developed in the previous section.

Consider the semidirect product Lie group $S:=G\,\circledS\,V$ associated to a right representation $\rho : G\rightarrow\operatorname{Aut}(V)$. The cotangent lift of the right translation is given by
\[
R^{T^*}_{(g,v)}(\alpha_f,(u,a))=(R^{T^*}_g(\alpha_f),v+\rho_g(u),\rho_{g^{-1}}^*(a))\in T^*_{(f,u)(g,v)}S.
\]
We modify this cotangent lifted action by an affine term of the form
\begin{equation}\label{C}
C_{(g,v)}(f,u):=(0_{fg},v+\rho_g(u),c(g)),
\end{equation}
for a group one-cocycle $c\in\mathcal{F}(G,V^*)$, that is, verifying the property $c(fg)=\rho_{g^{-1}}^*(c(f))+c(g)$. The resulting affine right action on $T^*S$ is therefore given by
\begin{align}\label{Psi}
\Psi_{(g,v)}(\alpha_f,(u,a)):&=R^{T^*}_{(g,v)}(\alpha_f,(u,a))+C_{(g,v)}(f,u)\nonumber\\
&=(R^{T^*}_g(\alpha_f),v+\rho_g(u),\rho_{g^{-1}}^*(a)+c(g))
\end{align}
This action is clearly of the form \eqref{affine_cot_lift}, and it is readily verified that property \eqref{cocycle_property} holds. This proves that $\Psi_{(g,v)}$ is a right action.

In the following lemma, we compute the one-form $\alpha\in\Omega^1_0(S)$ associated to $C$ and we show that it verifies the hypotheses of Theorem \ref{affine_LiePoisson_reduction}. Recall that $\alpha$ is defined by $\alpha(g,v):=C_{(g,v)}(e,0)$.

\begin{lemma}\label{dalpha} The one-form $\alpha \in \Omega^1_0(S) $ associated to the affine term \eqref{C} is given by
\begin{equation}
\label{group_alpha}
\alpha(g,v)(\xi_g,(v,u))=\langle c(g),u\rangle,
\end{equation}
for $(\xi_g,(v,u))\in T_{(g,v)}S$. Moreover $\mathbf{d}\alpha$ is $S$-invariant and its value at the identity is given by
\begin{equation}
\label{group_b}
\mathbf{d}\alpha(e,0)((\xi,u),(\eta,w))=\langle\mathbf{d}c(\xi),w\rangle-\langle \mathbf{d}c(\eta),u\rangle.
\end{equation}
The map $\phi : S\rightarrow\mathfrak{s}^*$ defined by
\[
\phi(g,v)=(\mathbf{d}c^T(v)-v\diamond c(g),-c(g)),
\]
verifies the property
\[
\mathbf{i}_{(\xi,u)^L}\mathbf{d}\alpha=\mathbf{d}\langle\phi,(\xi,u)\rangle,
\]
where $(\xi,u)^L\in\mathfrak{X}(S)$ is the left-invariant vector field induced by $(\xi,u)\in\mathfrak{s}$.
\end{lemma}

Using the equality $\sigma=-\phi$, we obtain that the bilinear form $\Sigma$, appearing in the formula of the affine orbit symplectic form and the affine Lie-Poisson bracket, is given by
\begin{align*}
\Sigma((\xi,u),\cdot)&=-T_{(e,0)}\sigma(\xi,u)
=-\left.\frac{d}{dt}\right|_{t=0}\left(tu\diamond c(\operatorname{exp}(t\xi))-\mathbf{d}c^T(tu),c(\operatorname{exp}(t\xi))\right)\\
&=(\mathbf{d}c^T(u),-\mathbf{d}c(\xi)),
\end{align*}
where $(\xi,u)\in\mathfrak{s}$

\medskip

\paragraph{The momentum map.} By item $\bf{iii}$ of Theorem \ref{affine_LiePoisson_reduction} and using the one-form $\alpha \in \Omega^1_0(S) $ given by \eqref{group_alpha}, we obtain that a momentum map for the right-action \eqref{Psi} is given by
\begin{align}\label{momentum_map}
\mathbf{J}_\alpha(\beta_f,(u,a))&=\mathbf{J}_R(t_\alpha(\beta_f,(u,a)))-\phi(f,u)\nonumber\\
&=(T^*L_f(\beta_f)+u\diamond a-\mathbf{d}c^T(u),a),
\end{align}
with nonequivariance one-cocycle
\begin{equation}\label{nonequivariance}
\sigma(f,u)=-\phi(f,u)=(u\diamond c(f)-\mathbf{d}c^T(u),c(f))\in\mathfrak{s}^*.
\end{equation}

\medskip

\paragraph{Poisson bracket and symplectic reduced spaces.} Using formula \eqref{affine_Lie_Poisson} and the expression of $\Sigma$, we obtain that the reduced Poisson bracket on $\mathfrak{s}^*$ is given by
\begin{align*}
\{f,g\}^+_\sigma(\mu,a)&=\left\langle\mu,\left[\frac{\delta f}{\delta\mu},\frac{\delta g}{\delta\mu}\right]\right\rangle+\left\langle a,\frac{\delta f}{\delta a}\frac{\delta g}{\delta\mu}-\frac{\delta g}{\delta a}\frac{\delta f}{\delta\mu}\right\rangle\\
&\qquad+\left\langle\mathbf{d}c\left(\frac{\delta f}{\delta\mu}\right),\frac{\delta g}{\delta a}\right\rangle-\left\langle\mathbf{d}c\left(\frac{\delta g}{\delta\mu}\right),\frac{\delta f}{\delta a}\right\rangle.
\end{align*}
By item $\bf{iv}$ of Theorem \ref{affine_LiePoisson_reduction}, the reduced space $\left(\mathbf{J}_\alpha^{-1}(\mu,a)/S^{\,\sigma}_{(\mu,a)},\Omega_{(\mu,a)}\right)$ is symplectically diffeomorphic to the affine coadjoint orbit $\left(\mathcal{O}^{\,\sigma}_{(\mu,a)},\omega^+_\sigma\right)$. More precisely, we have
\begin{align}\label{affine_coadj}
\mathcal{O}^{\,\sigma}_{(\mu,a)}&=\left\{\left.\operatorname{Ad}^*_{(g,u)}(\mu,a)+\sigma(g,u)\right| (g,u)\in S\right\}\nonumber\\
&=\left\{\left.\left(\operatorname{Ad}^*_g\mu+u\diamond(\rho_{g^{-1}}^*(a)+c(g))-\mathbf{d}c^T(u),\rho_{g^{-1}}^*(a)+c(g)\right)\right| (g,u)\in S \right\}.
\end{align}
The symplectic structure on $\mathcal{O}^{\,\sigma}_{(\mu,a)}$ is given by
\begin{align}\label{affineKKS}
\omega^+_\sigma(\lambda,b)&\left(\left(\operatorname{ad}^*_\xi\lambda+u\diamond b-\mathbf{d}c^T(u),b\xi+\mathbf{d}c(\xi)\right),\left(\operatorname{ad}^*_\eta\lambda+w\diamond b-\mathbf{d}c^T(w),b\eta+\mathbf{d}c(\eta)\right)\right)\nonumber\\
&=\left\langle \lambda,[\xi,\eta]\right\rangle+\left\langle b,u\eta-w\xi\right\rangle+\langle\mathbf{d}c(\eta),u\rangle-\langle\mathbf{d}c(\xi),w\rangle.
\end{align}

\medskip

\paragraph{Affine Lie-Poisson Hamiltonian reduction for semidirect products.} Consider a Hamiltonian function $H: T ^\ast G \times V ^\ast \rightarrow \mathbb{R}$ right-invariant under the $G$-action
\begin{equation}\label{affine_action}
(\alpha_h,a)\mapsto(R^{T^*}_g(\alpha_h),\theta_g(a))=(R^{T^*}_g(\alpha_h),\rho_{g^{-1}}^*(a)+c(g)).
\end{equation}
This $G $-action on $T ^\ast G \times V ^\ast$ is induced by the $S$-action \eqref{Psi}  on $T ^* S $. Note that we can think of this Hamiltonian $H:T^*G\times V^*\to\mathbb{R}$ as being the Poisson reduction of a $S$-invariant Hamiltonian $\overline{H}:T^*S\to\mathbb{R}$ by the normal subgroup $\{e\} \times V $ since $(T ^\ast S)/(\{e\} \times V)  \cong T ^\ast G \times V^\ast$. In particular, the
function $H_{a_0}: = H|_{T ^\ast G\times \{a_0\}}: T ^\ast G
\rightarrow \mathbb{R}$ is invariant under the induced action
of the isotropy subgroup $G_{a_0}^c$ of $a_0$ relative to the affine action $\theta$, for any $a_0 \in V ^\ast$. The following theorem is a generalization of Theorem \ref{LPSD} and is also a consequence of  the reduction
by stages method for nonequivariant momentum maps, together with the results obtained in section \ref{AMLPR} and at the beginning of the present section.

\begin{theorem}\label{ALPSD}
For $\alpha(t)\in T^*_{g(t)}G$ and
$\mu(t):=T^*R_{g(t)}(\alpha(t))\in\mathfrak{g}^*$, the following are
equivalent:
\begin{itemize}
\item[\bf{i}] $\alpha(t)$ satisfies Hamilton's equations for
$H_{a_0}$ on $T^*G$.
\item[\bf{ii}] The following affine Lie-Poisson equation holds on $\mathfrak{s}^*$:
\[
\frac{\partial}{\partial t}(\mu,a)=\left(-\operatorname{ad}^*_{\frac{\delta h}{\delta \mu}}\mu-\frac{\delta h}{\delta a}\diamond a+\mathbf{d}c^T\left(\frac{\delta h}{\delta a}\right),-a\frac{\delta h}{\delta \mu}-\mathbf{d}c\left(\frac{\delta h}{\delta \mu}\right)\right),\quad a(0)=a_0
\]
where $\mathfrak{s}$ is the semidirect product Lie algebra
$\mathfrak{s}=\mathfrak{g}\,\circledS\, V$. The associated Poisson bracket is the following affine Lie-Poisson bracket on the semidirect product Lie algebra $\mathfrak{s}^*$,
\begin{align*}
\{f,g\}_\sigma^+(\mu,a)&=\left\langle\mu,\left[\frac{\delta f}{\delta\mu},\frac{\delta g}{\delta\mu}\right]\right\rangle+\left\langle a,\frac{\delta f}{\delta a}\frac{\delta g}{\delta\mu}-\frac{\delta g}{\delta a}\frac{\delta f}{\delta\mu}\right\rangle\\
&\qquad+\left\langle\mathbf{d}c\left(\frac{\delta f}{\delta\mu}\right),\frac{\delta g}{\delta a}\right\rangle-\left\langle\mathbf{d}c\left(\frac{\delta g}{\delta\mu}\right),\frac{\delta f}{\delta a}\right\rangle.
\end{align*}
\end{itemize}
The evolution of the advected quantities is given by $a(t)=\theta_{g(t)^{-1}}(a_0)$.
\end{theorem}


\section{Hamiltonian Approach to Continuum Theories of Complex Fluids}\label{Hamiltonian_PCF}

Recall that in the case of the motion of a fluid on an orientable manifold $\mathcal{D}$, the configuration space is the group $G=\operatorname{Diff}(\mathcal{D})$ of all diffeomorphisms of $\mathcal{D}$. In the case of incompressible fluids, one chooses the subgroup $\operatorname{Diff}_{\rm vol}(\mathcal{D})$ of all volume preserving diffeomorphisms, with respect to a fixed volume form on $\mathcal{D}$. Besides the diffeomorphism group, the other basic object is the vector space $V^*$ of advected quantities on which $G$ acts by representations. Typical advected quantities are for example the \textit{mass density}, the \textit{entropy density} or the \textit{magnetic field}. One can obtain the fluid equations by choosing the appropriate Hamiltonian function and by applying the semidirect Lie-Poisson reduction process (Theorem \ref{LPSD}), see \cite{MaRaWe1984} and \cite{HoMaRa1998}.

The goal of this section is to extend these formulations to the case of complex fluids. At the reduced level, the affine Lie-Poisson equations for complex fluids are given in \cite{Ho2002} (equation (3.44)). The two key observations we make regarding these equations are the following. First, they suggest that the configuration manifold $\operatorname{Diff}(\mathcal{D})$ has to be enlarged to a bigger group $G$ in order to contain variables involving the Lie group $\mathcal{O}$ of order parameters. Second they suggest that there is a new advected quantity on which the group $G$ acts by affine representation. Making use of these two observations, we construct below the appropriate configuration space and the appropriate affine action for the dynamics of complex fluids. By using the general process of affine Lie-Poisson reduction developed before (Theorem \ref{ALPSD}), we get (a generalization of) the equations given in \cite{Ho2002}.

Here and in all examples that follow, there are fields different from the velocity field for which we shall never specify the boundary conditions.
We make the general assumption, valid throughout the paper, that all integrations by parts have vanishing boundary terms, or that the problem has periodic boundary conditions (in which case $\mathcal{D}$ is a boundaryless three dimensional manifold). Of course if one would try to get an analytically rigorous result, the boundary conditions for all fields need to be carefully specified.

\paragraph{The configuration manifold.}

Consider a finite dimensional Lie group $\mathcal{O}$. In applications $\mathcal{O}$ will be called the \textit{order parameter Lie group}. Recall that in the case of the motion of a fluid on an orientable manifold $\mathcal{D}$, the configuration space is the group $G=\operatorname{Diff}(\mathcal{D})$ of all diffeomorphisms of $\mathcal{D}$. In the case of complex fluids, the basic idea is to enlarge this group to the semidirect product of groups $G=\operatorname{Diff}(\mathcal{D})\,\circledS\,\mathcal{F}(\mathcal{D},\mathcal{O})$. Here $\mathcal{F}(\mathcal{D},\mathcal{O})$ denotes the group of all mappings $\chi$ defined on $\mathcal{D}$ with values in the Lie group $\mathcal{O}$ of order parameters. The diffeomorphism group acts on $\mathcal{F}(\mathcal{D},\mathcal{O})$ via the \textit{right\/} action
\[
(\eta,\chi)\in\operatorname{Diff}(\mathcal{D})\times\mathcal{F}(\mathcal{D},\mathcal{O})\mapsto\chi\circ\eta\in\mathcal{F}(\mathcal{D},\mathcal{O}).
\]
Therefore, the group multiplication is given by
\[
(\eta,\chi)(\varphi,\psi)=(\eta\circ\varphi,(\chi\circ\varphi)\psi).
\]

Recall that the tangent space to $\operatorname{Diff}(\mathcal{D})$ at $\eta$ is
\[
T_\eta\operatorname{Diff}(\mathcal{D})=\{\mathbf{u}_\eta:\mathcal{D}\rightarrow T\mathcal{D}\mid \mathbf{u}_\eta(x)\in T_{\eta(x)}\mathcal{D}\},
\]
the tangent space to $\mathcal{F}(\mathcal{D},\mathcal{O})$ at $\chi$ is
\[
T_\chi\mathcal{F}(\mathcal{D},\mathcal{O})=\{\nu_\chi:\mathcal{D}\rightarrow T\mathcal{O}\mid \nu_\chi(x)\in T_{\chi(x)}\mathcal{O}\}.
\]
A direct computation shows that the tangent map of right translation is 
\[
TR_{(\varphi,\psi)}(\mathbf{u}_\eta,\nu_\chi)=(\mathbf{u}_\eta\circ\varphi,TR_\psi(\nu_\chi\circ\varphi)).
\]

For simplicity we fix a volume form $\mu$ on $\mathcal{D}$. Therefore we can identify the cotangent space $T_\eta^*\operatorname{Diff}(\mathcal{D})$ with a space of one-forms over $\eta$, that is,
\[
T_\eta^*\operatorname{Diff}(\mathcal{D})=\{\mathbf{m}_\eta:\mathcal{D}\rightarrow T^*\mathcal{D}\mid\mathbf{m}_\eta(x)\in T^*_{\eta(x)}\mathcal{D}\}.
\]
The cotangent space of $\mathcal{F}(\mathcal{D},\mathcal{O})$ at $\chi$ is naturally given by
\[
T^*_\chi\mathcal{F}(\mathcal{D},\mathcal{O})=\{\kappa_\chi:\mathcal{D}\rightarrow T^*\mathcal{O}\mid \kappa_\chi(x) \in  T^*_{\chi(x)} \mathcal{O}\}.
\]
Using these identifications, the cotangent lift of right translation is given by
\[
R^{T^*}_{(\varphi,\psi)}(\mathbf{m}_\eta,\kappa_\chi)=J(\varphi)(\mathbf{m}_\eta\circ\varphi,T^*R_{\psi^{-1}}(\kappa_\chi\circ\varphi)),
\]
where $J(\varphi) $ is the Jacobian determinant of the diffeomorphism $\varphi$.
The Lie algebra $\mathfrak{g}$ of the semidirect product group is
\[
\mathfrak{g}=\mathfrak{X}(\mathcal{D})\,\circledS\,\mathcal{F}(\mathcal{D},\mathfrak{o}),
\]
and the Lie bracket is computed to be
\[
\operatorname{ad}_{(\mathbf{u},\nu)}(\mathbf{v},\zeta)=(\operatorname{ad}_{\mathbf{u}}\mathbf{v},\operatorname{ad}_{\nu}\zeta+\mathbf{d}\nu\cdot\mathbf{v}-\mathbf{d}\zeta\cdot\mathbf{u}),
\]
where $\operatorname{ad}_{\mathbf{u}}\mathbf{v}=-[\mathbf{u},\mathbf{v}]$, $\operatorname{ad}_{\nu}\zeta\in\mathcal{F}(\mathcal{D},\mathfrak{o})$ is given by $\operatorname{ad}_{\nu}\zeta(x):=\operatorname{ad}_{\nu(x)}\zeta(x)$, and $\mathbf{d}\nu\cdot\mathbf{v}\in\mathcal{F}(\mathcal{D},\mathfrak{o})$ is given by $\mathbf{d}\nu\cdot\mathbf{v}(x):=\mathbf{d}\nu(x)(\mathbf{v}(x))$.

Using the previous identification of cotangent spaces, the dual Lie algebra $\mathfrak{g}^*$ can be identified with $\Omega^1(\mathcal{D})\,\circledS\,\mathcal{F}(\mathcal{D},\mathfrak{o}^*)$ through the pairing
\[
\langle(\mathbf{m},\kappa),(\mathbf{u},\nu)\rangle=\int_\mathcal{D}\left(\mathbf{m}\cdot\mathbf{u}+\kappa\cdot\nu\right)\mu.
\]
The dual map to $\operatorname{ad}_{(\mathbf{u},\nu)}$ is 
\begin{equation}\label{ad*}
\operatorname{ad}^*_{(\mathbf{u},\nu)}(\mathbf{m},\kappa)=\big({\boldsymbol{\pounds}}_\mathbf{u}\mathbf{m}+(\operatorname{div}\mathbf{u})\mathbf{m}+\kappa\cdot\mathbf{d}\nu,\operatorname{ad}^*_\nu\kappa+\operatorname{div}(\mathbf{u}\kappa)\big).
\end{equation}
This formula needs some explanation. The symbol $\kappa\cdot\mathbf{d}\nu\in\Omega^1(\mathcal{D})$ denotes the 
one-form defined by 
\[
\kappa\cdot\mathbf{d}\nu(v_x):=\kappa(x)(\mathbf{d}\nu(v_x)).
\]
The expression $\mathbf{u}\kappa$ denotes the $1$-contravariant tensor field with values in $\mathfrak{o}^*$ defined by
\[
\mathbf{u}\kappa(\alpha_x):=\alpha_x(\mathbf{u}(x))\kappa(x)\in\mathfrak{o}^*.
\]
Since $\mathbf{u}\kappa $ is a generalization of the notion of a vector field, we denote by $\mathfrak{X}(\mathcal{D},\mathfrak{o}^*)$ the space of all $1$-contravariant tensor fields with values in $\mathfrak{o}^*$. In \eqref{ad*}, $\operatorname{div}(\mathbf{u})$ denotes the divergence of the vector field $\mathbf{u}$ with respect to the fixed volume form $\mu$. Recall that it is defined by the condition
\[
(\operatorname{div}\mathbf{u})\mu={\boldsymbol{\pounds}}_\mathbf{u}\mu.
\]
This operator can be naturally extended to the space $\mathfrak{X}(\mathcal{D},\mathfrak{o}^*)$ as follows. For $w\in\mathfrak{X}(\mathcal{D},\mathfrak{o}^*)$ we write $w=w_a\varepsilon^a$ where $(\varepsilon^a)$ is a basis of $\mathfrak{o}^*$ and $w_a\in\mathfrak{X}(\mathcal{D})$. We define $\operatorname{div}:\mathfrak{X}(\mathcal{D},\mathfrak{o}^*)\rightarrow\mathcal{F}(\mathcal{D},\mathfrak{o}^*)$ by the equality
\[
\operatorname{div} w:=(\operatorname{div}w_a)\varepsilon^a.
\]
Note that for $w=\mathbf{u}\kappa$ we have  
\[
\operatorname{div}(\mathbf{u}\kappa)=\textbf{d}\kappa\cdot\mathbf{u}+(\operatorname{div}\mathbf{u})\kappa.
\]

\paragraph{The space of advected quantities.}
In physical applications, the affine representation space $V^*$ of $G=\operatorname{Diff}(\mathcal{D})\,\circledS\,\mathcal{F}(\mathcal{D},\mathcal{O})$ is a direct product $V_1^*\oplus V^*_2$, where $V_i^*$ are subspaces of the space of all tensor fields on $\mathcal{D}$ (possibly with values in a vector space). Moreover:
\begin{itemize}
\item $V_1^*$ is only acted upon by the component $\operatorname{Diff}(\mathcal{D})$ of $G$.
\item The action of $G$ on $V_2^*$ is affine, with the restriction that the affine term only depends on the second component $\mathcal{F}(\mathcal{D},\mathcal{O})$ of $G$.
\end{itemize}

In this way, we obtain the affine representation 
\begin{equation}\label{affine_representation}
(a,\gamma)\in V_1^*\oplus V^*_2\mapsto (a\eta,\gamma(\eta,\chi)+C(\chi))\in V_1^*\oplus V^*_2,
\end{equation}
where $\gamma(\eta,\chi)$ denotes the representation of $(\eta,\chi)\in G $ on  $\gamma\in V^*_2$, and $C\in\mathcal{F}(\mathcal{F}(\mathcal{D},\mathcal{O}),V^*_2)$ satisfies the identity
\begin{equation}\label{cocyclecondition}
C((\chi\circ\varphi)\psi)=C(\chi)(\varphi,\psi)+C(\psi).
\end{equation}
Note that this is equivalent to say that the representation $\rho$ and the affine term $c$ of the previous section have the particular form
\[
\rho^*_{(\eta,\chi)^{-1}}(a,\gamma)=(a\eta,\gamma(\eta,\chi))\quad\text{and}\quad c(\eta,\chi)=(0,C(\chi)).
\]

The infinitesimal action of $(\mathbf{u}, \nu) \in \mathfrak{g}$ on $\gamma\in V_2^*$ induced by the representation of $G$ on $V_2^*$ is
\begin{align*}
\gamma(\mathbf{u},\mathbf{\nu}):&=\left.\frac{d}{dt}\right|_{t=0}\gamma(\operatorname{exp}(t\mathbf{u}),\operatorname{exp}(t\nu))=\left.\frac{d}{dt}\right|_{t=0}\gamma(\operatorname{exp}(t\mathbf{u}),e)(e,\operatorname{exp}(t\nu))\\
&=\left.\frac{d}{dt}\right|_{t=0}\gamma(\operatorname{exp}(t\mathbf{u}),e)+\left.\frac{d}{dt}\right|_{t=0}\gamma(e,\operatorname{exp}(t\nu))=:\gamma\mathbf{u}+\gamma\nu.
\end{align*}
Therefore, for $(v,w)\in V_1\oplus V_2$ we have
\[
(v,w)\diamond (a,\gamma)=(v\diamond a+w\diamond_1\gamma,w\diamond_2\gamma),
\]
where $\diamond_1$ and $\diamond_2$ are associated to the induced representations of the first and second component of $G$ on $V^*_2$. On the right hand side, the diamond operation $\diamond$ is associated to the representation of $\operatorname{Diff}(\mathcal{D})$ on $V_1^*$. The space $V_1^*$ is naturally the dual of some space $V_1$ of tensor fields on $\mathcal{D}$. For example the $(p,q)$ tensor fields are naturally in duality with the $(q,p)$ tensor fields. For $a\in V_1^*$ and $v\in V_1$, the duality pairing is given by
\[
\langle a,v\rangle=\int_\mathcal{D}(a\cdot v)\mu,
\]
where $\cdot$ denotes the contraction of tensor fields.

Since the affine cocycle has the particular form $c(\eta,\chi)=(0,C(\chi))$, we obtain that
\[
\mathbf{d}c^T(v,w)=(0,\mathbf{d}C^T(w)).
\]

By Theorem \ref{ALPSD}, we obtain that the associated affine Lie-Poisson bracket is given by
\begin{align*}
\{f,g\}_\sigma^+(\mathbf{m},\kappa,a,\gamma)&=\int_\mathcal{D}\mathbf{m}\cdot\left[\frac{\delta f}{\delta\mathbf{m}},\frac{\delta g}{\delta\mathbf{m}}\right]\mu\\
&\quad+\int_\mathcal{D}\kappa\cdot\left(\operatorname{ad}_{\frac{\delta f}{\delta\kappa}}\frac{\delta g}{\delta\kappa}+\mathbf{d}\frac{\delta f}{\delta\kappa}\cdot\frac{\delta g}{\delta\mathbf{m}}-\mathbf{d}\frac{\delta g}{\delta\kappa}\cdot\frac{\delta f}{\delta\mathbf{m}}\right)\mu\\
&\quad+\int_\mathcal{D}a\cdot\left(\frac{\delta f}{\delta a}\frac{\delta g}{\delta\mathbf{m}}-\frac{\delta g}{\delta a}\frac{\delta f}{\delta\mathbf{m}}\right)\\
&\quad+\int_\mathcal{D}\gamma\cdot\left(\frac{\delta f}{\delta\gamma}\frac{\delta g}{\delta\mathbf{m}}+\frac{\delta f}{\delta\gamma}\frac{\delta g}{\delta\kappa}-\frac{\delta g}{\delta\gamma}\frac{\delta f}{\delta\mathbf{m}}-\frac{\delta g}{\delta\gamma}\frac{\delta f}{\delta\kappa}\right)\mu\\
&\quad+\int_\mathcal{D}\left(\mathbf{d}C\left(\frac{\delta f}{\delta\kappa}\right)\cdot\frac{\delta g}{\delta\gamma}-\mathbf{d}C\left(\frac{\delta g}{\delta\kappa}\right)\cdot\frac{\delta f}{\delta\gamma}\right)\mu.
\end{align*}
The first four terms give the Lie-Poisson bracket on the dual Lie algebra
\[
\left( \left[\mathfrak{X}( \mathcal{D}) \,\circledS\, \mathcal{F}( \mathcal{D}, \mathfrak{o}) \right] \,\circledS\, \left[V _1\oplus V _2 \right] \right)^\ast \cong 
\Omega^1(\mathcal{D}) \times \mathcal{F}(\mathcal{D},\mathfrak{o}^*) \times V^*_1\ \times V_2^*.
\]
The last term is due to the presence of the affine term $C$ in the representation. Since $C$ depends only on the group $\mathcal{F}(\mathcal{D},\mathcal{O})$, this term does not involve the functional derivatives with respect to $\mathbf{m}$.

The symplectic leaves of this bracket are the affine coadjoint orbits in the dual Lie algebra
$\left( \left[\mathfrak{X}( \mathcal{D}) \,\circledS\, \mathcal{F}( \mathcal{D}, \mathfrak{o}) \right] \,\circledS\, \left[V _1\oplus V _2 \right] \right)^\ast $. 
The expression of the tangent spaces and of the affine orbit symplectic forms involves the bilinear form $\Sigma$ which is defined in this case on $\left[\mathfrak{X}(\mathcal{D})\,\circledS\,\mathcal{F}(\mathcal{D},\mathfrak{o})\right]\,\circledS\,\left[V_1\,\oplus\,V_2\right]$ by
\[
\Sigma((\mathbf{u}_1,\nu_1,v_1,w_1),(\mathbf{u}_2,\nu_2,v_2,w_2))=\mathbf{d}C(\nu_2)\cdot w_1-\mathbf{d}C(\nu_1)\cdot w_2
\] 

For a Hamiltonian $h=h(\mathbf{m},\kappa,a,\gamma):\Omega^1(\mathcal{D}) \times \mathcal{F}(\mathcal{D},\mathfrak{o}^*) \times V^*_1\ \times V_2^* \to\mathbb{R}$, the affine Lie-Poisson equations of Theorem \ref{ALPSD} become
\begin{equation}\label{ALP_PCF}
\left\{
\begin{array}{ll}
\vspace{0.1cm}\displaystyle\frac{\partial}{\partial t}\mathbf{m}=-{\boldsymbol{\pounds}}_{\frac{\delta h}{\delta \mathbf{m}}}\mathbf{m}-\operatorname{div}\left(\frac{\delta h}{\delta \mathbf{m}}\right)\mathbf{m}-\kappa\cdot\mathbf{d}\frac{\delta h}{\delta \kappa}-\frac{\delta h}{\delta a}\diamond a-\frac{\delta h}{\delta \gamma}\diamond_1\gamma\\
\vspace{0.1cm}\displaystyle\frac{\partial}{\partial t}\kappa=-\operatorname{ad}^*_{\frac{\delta h}{\delta \kappa}}\kappa-\operatorname{div}\left(\frac{\delta h}{\delta \mathbf{m}}\kappa\right)-\frac{\delta h}{\delta\gamma}\diamond_2\gamma+\mathbf{d}C^T\left(\frac{\delta h}{\delta\gamma}\right)\\
\vspace{0.1cm}\displaystyle\frac{\partial}{\partial t}a=-a\frac{\delta h}{\delta \mathbf{m}}\\
\vspace{0.1cm}\displaystyle\frac{\partial}{\partial t}\gamma=-\gamma\frac{\delta h}{\delta \mathbf{m}}-\gamma\frac{\delta h}{\delta \kappa}-\mathbf{d}C\left(\frac{\delta h}{\delta \kappa}\right).
\end{array}
\right.
\end{equation}

Using formula \eqref{momentum_map}, the momentum map of the affine right action of the semidirect product $[\operatorname{Diff}(\mathcal{D})\,\circledS\,\mathcal{F}(\mathcal{D},\mathcal{O})]\,\circledS\,[V_1\oplus V_2]$ on its cotangent bundle is computed to be
\begin{align*}
&\mathbf{J}_\alpha(\mathbf{m}_\eta,\kappa_\chi,(v,w),(a,\gamma))\\
&=\left(T^*\eta\circ\mathbf{m}_\eta+T^*\chi\circ\kappa_\chi+v\diamond a+w\diamond_1\gamma,T^*L_\chi\circ\kappa_\chi+w\diamond_2\gamma-\mathbf{d}C^T(w),(a,\gamma)\right).
\end{align*}

By the general theory, the nonequivariance cocycle of $\mathbf{J}_\alpha$ is given by $\sigma=-\phi$, where $\phi$ is computed to be
\[
\phi : [\operatorname{Diff}(\mathcal{D})\,\circledS\,\mathcal{F}(\mathcal{D},\mathcal{O})]\,\circledS\,[V_1\oplus V_2]\to\Omega^1(\mathcal{D}) \times \mathcal{F}(\mathcal{D},\mathfrak{o}^*) \times V^*_1\ \times V_2^*,
\]
\begin{align*}
\phi(\eta,\chi,v,w)&=(\mathbf{d}c^T(v,w)-(v,w)\diamond c(\eta,\chi),-c(\eta,\chi))\\
&=(-w\diamond_1 C(\chi),\mathbf{d}C^T(w)-w\diamond_2 C(\chi),0,-C(\chi)).
\end{align*}

\paragraph{Basic example:} Take $V_2^*:=\Omega^1(\mathcal{D},\mathfrak{o})$, the space of all one-forms on $\mathcal{D}$ with values in $\mathfrak{o}$. This space is naturally the dual of the space $V_2=\mathfrak{X}(\mathcal{D},\mathfrak{o}^*)$ of contravariant tensor fields with values in $\mathfrak{0}^*$, the duality pairing being given, for $\gamma\in\Omega^1(\mathcal{D},\mathfrak{o})$ and $w\in\mathfrak{X}(\mathcal{D},\mathfrak{o}^*)$, by
\[
\langle\gamma,w\rangle:=\int_\mathcal{D}(\gamma\cdot w)\mu,
\]
where $\gamma\cdot w$ denotes the contraction of tensors.

We consider for \eqref{affine_representation} the affine representation defined by
\begin{equation}\label{affine_representation_gamma}
(a,\gamma)\mapsto(a\eta,\operatorname{Ad}_{\chi^{-1}}\eta^*\gamma+\chi^{-1}T\chi).
\end{equation}
One can check that $\gamma(\eta,\chi):=\operatorname{Ad}_{\chi^{-1}}\eta^*\gamma$ is a right representation of $G$ on $V_2^*$ and that $C(\chi)=\chi^{-1}T\chi$ verifies the condition \eqref{cocyclecondition}. The one-form $\gamma \in \Omega^1( \mathcal{D}, \mathfrak{o}) $  can be considered as a connection one-form on the trivial principal $\mathcal{O}$-bundle $\mathcal{O} \times \mathcal{D} \rightarrow \mathcal{D}$. The covariant differential associated to this principal connection will be denoted by $\mathbf{d}^ \gamma$. Therefore, for a function $\nu\in\mathcal{F}(\mathcal{D},\mathfrak{o})$, we have 
\[
\mathbf{d}^\gamma\nu(\mathbf{v}):=\mathbf{d}\nu(\mathbf{v})+[\gamma(\mathbf{v}),\nu].
\]
The covariant divergence of $w\in\mathfrak{X}(\mathcal{D},\mathfrak{o}^*)$ is the function
\[
\operatorname{div}^\gamma w:=\operatorname{div}w-\operatorname{Tr}(\operatorname{ad}^*_\gamma w)\in\mathcal{F}(\mathcal{D},\mathfrak{o}^*),
\]
defined as minus the adjoint of the covariant differential, that is, 
\begin{equation}
\label{divergence_formula}
\int_\mathcal{D}\left(\mathbf{d}^\gamma\nu\cdot w\right)\mu=-\int_\mathcal{D}\left(\nu\cdot\operatorname{div}^\gamma w\right)\mu
\end{equation}
for all $\nu \in \mathcal{F}( \mathcal{D}, \mathfrak{o}) $. More generally, the space $\Omega^k(\mathcal{D},\mathfrak{o})$ is, in a natural way, dual to the space $\Omega_k(\mathcal{D},\mathfrak{o}^*)$ of $k$-contravariant skew symmetric tensor fields with values in $\mathfrak{o}^*$. The duality pairing is given by contraction and integration with respect to the fixed volume form $\mu$. Therefore, we can define the divergence operators, $\operatorname{div}, \operatorname{div} ^ \gamma :\Omega_k(\mathcal{D},\mathfrak{o}^*)\rightarrow\Omega_{k-1}(\mathcal{D},\mathfrak{o}^*)$, to be minus the adjoint of the exterior derivatives $\mathbf{d}$ and $\mathbf{d}^\gamma$, respectively. Note that we have $\Omega_1(\mathcal{D},\mathfrak{o}^*)=\mathfrak{X}(\mathcal{D},\mathfrak{o}^*)$ and $\Omega_0(\mathcal{D},\mathfrak{o}^*)=\mathcal{F}(\mathcal{D},\mathfrak{o}^*)$.

For the particular case of the affine action \eqref{affine_representation_gamma}, we have
\[
\gamma\mathbf{u}={\boldsymbol{\pounds}}_\mathbf{u}\gamma=\mathbf{d}^\gamma(\gamma(\mathbf{u}))+\mathbf{i}_\mathbf{u}\mathbf{d}^\gamma\gamma,
\]
\[
\quad\gamma\nu=-\operatorname{ad}_\nu\gamma,\quad\mathbf{d}C(\nu)=\mathbf{d}\nu,\quad\text{and}\quad\mathbf{d}C^T(w)=-\operatorname{div}(w).
\]
The diamond operations are given by
\[
w\diamond_1\gamma=(\operatorname{div}^\gamma w)\cdot\gamma-w\cdot\mathbf{i}_{\_\,}\mathbf{d}^\gamma\gamma\quad\text{and}\quad w\diamond_2\gamma=-\operatorname{Tr}\left(\operatorname{ad}^*_\gamma w\right).
\]
The affine Lie-Poisson equations \eqref{ALP_PCF} become
\begin{equation}\label{ALP_PerfectComplexFluid}
\left\{
\begin{array}{ll}
\vspace{0.1cm}\displaystyle\frac{\partial}{\partial t}\mathbf{m}=-{\boldsymbol{\pounds}}_{\frac{\delta h}{\delta \mathbf{m}}}\mathbf{m}-\operatorname{div}\left(\frac{\delta h}{\delta \mathbf{m}}\right)\mathbf{m}-\kappa\cdot\mathbf{d}\frac{\delta h}{\delta \kappa}-\frac{\delta h}{\delta a}\diamond a \\
\vspace{0.1cm} \qquad \qquad \; \displaystyle 
-\left(\operatorname{div}^\gamma \frac{\delta h}{\delta \gamma}\right)\gamma+\frac{\delta h}{\delta \gamma}\cdot\mathbf{i}_{\_\,}\mathbf{d}^\gamma\gamma\\
\vspace{0.1cm}\displaystyle\frac{\partial}{\partial t}\kappa=-\operatorname{ad}^*_{\frac{\delta h}{\delta \kappa}}\kappa-\operatorname{div}\left(\frac{\delta h}{\delta \mathbf{m}}\kappa\right)-\operatorname{div}^\gamma\frac{\delta h}{\delta\gamma}\\
\vspace{0.1cm}\displaystyle\frac{\partial}{\partial t}a=-a\frac{\delta h}{\delta \mathbf{m}}\\
\vspace{0.1cm}\displaystyle\frac{\partial}{\partial t}\gamma=-\mathbf{d}^\gamma\left(\gamma\left(\frac{\delta h}{\delta \mathbf{m}}\right)\right)-\mathbf{i}_\frac{\delta h}{\delta \mathbf{m}}\mathbf{d}^\gamma\gamma-\mathbf{d}^\gamma\frac{\delta h}{\delta \kappa}\\
\end{array}
\right.
\end{equation}

So we recover, by a reduction from a canonical situation, the equations (3.44) of \cite{Ho2002}, up to sign conventions, as well as their Hamiltonian structure. In matrix notation and with respect to local coordinates we have
\begin{equation}\label{hamiltonian_matrix}
\left[\begin{array}{c}
\dot m_i\\
\dot\kappa_a\\
\dot a\\
\dot \gamma^a_i\end{array}\right]\!
=\!-\!\left[\begin{array}{cccc}
m_k\partial_i+\partial_km_i     &\kappa_b\partial_i&(\square\diamond a)_i&\partial_j\gamma^b_i-\gamma^b_{j,i}\\
\partial_k\kappa_a&\kappa_cC^c_{ba}&0&\delta^b_a\partial_j-C^b_{ca}\gamma^c_j\\
a\square\partial_k&0&0&0\\
\gamma_k^a\partial_i+\gamma_{i,k}^a&\delta_b^a\partial_i+C^a_{cb}\gamma_i^c&0&0\end{array}\right]\!\displaystyle\left[\begin{array}{c}
(\delta h/\delta  m)^k\\
(\delta h/\delta\kappa)^b\\
\delta h/\delta a\\
(\delta h/\delta \gamma)_b^j\end{array}\right]
\end{equation}
This matrix appears in \cite{HoKu1988} (formula (2.26a)) and in \cite{Ho2002} (formula (3.46)) as the common Hamiltonian structure for various hydrodynamical systems. See \cite{CeMaRa2003} for another derivation of this Hamiltonian structure based on reduction.

The associated affine Lie-Poisson bracket is
\begin{align}\label{ALPB}
\{f,g\}_\sigma^+(\mathbf{m},\kappa,a,\gamma)&=\int_\mathcal{D}\mathbf{m}\cdot\left[\frac{\delta f}{\delta\mathbf{m}},\frac{\delta g}{\delta\mathbf{m}}\right]\mu\\
&\quad+\int_\mathcal{D}\kappa\cdot\left(\operatorname{ad}_{\frac{\delta f}{\delta\kappa}}\frac{\delta g}{\delta\kappa}+\mathbf{d}\frac{\delta f}{\delta\kappa}\cdot\frac{\delta g}{\delta\mathbf{m}}-\mathbf{d}\frac{\delta g}{\delta\kappa}\cdot\frac{\delta f}{\delta\mathbf{m}}\right)\mu\nonumber\\
&\quad+\int_\mathcal{D}a\cdot\left(\frac{\delta f}{\delta a}\frac{\delta g}{\delta\mathbf{m}}-\frac{\delta g}{\delta a}\frac{\delta f}{\delta\mathbf{m}}\right)\mu\nonumber\\
&\quad+\int_\mathcal{D}\left[\left(\mathbf{d}^\gamma\frac{\delta f}{\delta\kappa}+{\boldsymbol{\pounds}}_{\frac{\delta f}{\delta\mathbf{m}}}\gamma\right)\cdot\frac{\delta g}{\delta\gamma}-\left(\mathbf{d}^\gamma\frac{\delta g}{\delta\kappa}+{\boldsymbol{\pounds}}_{\frac{\delta g}{\delta\mathbf{m}}}\gamma\right)\cdot\frac{\delta f}{\delta\gamma}\right]\mu\nonumber.
\end{align}

The momentum map is computed to be
\begin{align*}
&\mathbf{J}_\alpha(\mathbf{m}_\eta,\kappa_\chi,(v,w),(a,\gamma))\\
&=\left(T^*\eta\circ\mathbf{m}_\eta+T^*\chi\circ\kappa_\chi+v\diamond a+(\operatorname{div}^\gamma w)\cdot\gamma-w\cdot\mathbf{i}_{\_\,}\mathbf{d}^\gamma\gamma,T^*L_\chi\circ\kappa_\chi+\operatorname{div}^\gamma w,(a,\gamma)\right).
\end{align*}


\section{The Circulation Theorems}

The Kelvin-Noether theorem is a version of the Noether theorem that holds for
solutions of the Euler-Poincar\'e equations. For example, an application of this theorem to
the compressible adiabatic fluid gives the Kelvin
circulation theorem
\[
\frac{d}{dt}\oint_{\gamma_t}\mathbf{u}^\flat=\oint_{\gamma_t}T\textbf{d}s,
\]
where $\gamma_t\subset \mathcal{D}$ is a closed curve which moves with the fluid velocity
$\mathbf{u}$, $T$  is the \textit{temperature}, and $s$ denotes the \textit{specific entropy}. The Kelvin-Noether theorem associated to Euler-Poincar\'e reduction for semidirect product is presented in \cite{HoMaRa1998}. We now adapt this result to the case of affine Lie-Poisson reduction.

\paragraph{Kelvin-Noether Theorem.} Let
$\mathcal{C}$ be a manifold
on which $G$ acts on the left and suppose we have an equivariant map
$\mathcal{K}:\mathcal{C}\times V^*\rightarrow\mathfrak{g}^{**}$, that is, for
all $g\in G, a\in V^*, c\in\mathcal{C}$, we have
\[
\langle\mathcal{K}(gc,\theta_g(a)),\mu\rangle=\langle\mathcal{K}(c,a),\operatorname{Ad}_g^*\mu\rangle,
\]
where $gc$ denotes the action of $G$ on $\mathcal{C}$, and $\theta_g$ is the affine action of $G$ on $V^*$. Consider the map $J:\mathcal{C}\times\mathfrak{g}^*\times V^*\rightarrow\mathbb{R}$ defined by
\[
J(c,\mu,a):=\langle\mathcal{K}(c,a),\mu\rangle.
\]
We have the following result.

\begin{theorem}\label{KN_Hamiltonian}
Fixing $c_0\in\mathcal{C}$, let $\mu(t),a(t)$ satisfy the
affine Lie-Poisson equations of Theorem \ref{ALPSD} and define $g(t)$ to be the solution of
\[
\dot{g}(t)=TR_{g(t)}\left(\frac{\delta h}{\delta\mu}\right),\quad g(0)=e.
\]
Let $c(t)=g(t)c_0$ and
$J(t):=J(c(t),\mu(t),a(t))$. Then
\[
\frac{d}{dt}J(t)=\left\langle\mathcal{K}(c(t),a(t)),-\frac{\delta h}{\delta
a}\diamond a+\mathbf{d}c^T\left(\frac{\delta h}{\delta
a}\right)\right\rangle.
\]
\end{theorem}

In the case of dynamics on the group $G=\operatorname{Diff}(\mathcal{D})$, the standard choice for the equivariant map $\mathcal{K}$ is
\begin{equation}\label{K}
\langle\mathcal{K}(c,a),\mathbf{m}\rangle:=\oint_c\frac{1}{\rho}\mathbf{m},
\end{equation}
where $c\in\mathcal{C}=\operatorname{Emb}(S^1,\mathcal{D})$, the manifold of all embeddings of the circle $S^1$ in $\mathcal{D}$, $\mathbf{m}\in\Omega^1(\mathcal{D})$, and $\rho$ is advected as $(J\eta)(\rho\circ\eta)$. There is a generalization of this map in the case of the group $G=\operatorname{Diff}(\mathcal{D})\,\circledS\,\mathcal{F}(\mathcal{D},\mathcal{O})$, see \S7 in \cite{GBRa2008a}. Therefore, Theorem \ref{KN_Hamiltonian} can be applied in the case of the affine Lie-Poisson equations \eqref{ALP_PCF}. Nevertheless we shall not use this point of view here and we apply the Kelvin-Noether theorem only to the first component of the group $G$, namely the group $\operatorname{Diff}(\mathcal{D})$, and we obtain the following result.

\begin{theorem} Consider the affine Lie-Poisson equations for complex fluids \eqref{ALP_PCF}. Suppose that one of the linear advected variables, say $\rho$, is advected as $(J\eta)(\rho\circ\eta)$. Then, using the map 
\eqref{K}, we have
\[
\frac{d}{dt}\oint_{c_t}\frac{1}{\rho}\mathbf{m}=\oint_{c_t}\frac{1}{\rho}\left(-\kappa\cdot\mathbf{d}\frac{\delta h}{\delta \kappa}-\frac{\delta h}{\delta a}\diamond a-\frac{\delta h}{\delta \gamma}\diamond_1 \gamma\right),
\]
where $c_t$ is a loop in $\mathcal{D}$ which moves with the fluid velocity $\mathbf{u}$, defined be the equality
\[
\mathbf{u}:=\frac{\delta h}{\delta\mathbf{m}}.
\]
\end{theorem}

\paragraph{$\gamma$-circulation.} The $\gamma$-circulation is associated to the equation
\[
\frac{\partial}{\partial t}\gamma+\boldsymbol{\pounds}_\mathbf{u}\gamma=-\mathbf{d}\nu+\operatorname{ad}_\nu\gamma.
\]
Let $\eta_t$ be the flow of the vector field $\mathbf{u}$, let $c_0$ be a loop in $\mathcal{D}$ and let $c_t:=\eta_t \circ c_0$. Then, by change of variables, we have
\[
\frac{d}{dt}\oint_{c_t}\gamma=\frac{d}{dt}\oint_{c_0}\eta_t^*\gamma=\oint_{c_0}\eta_t^*(\dot\gamma+\boldsymbol{\pounds}_\mathbf{u}\gamma)=\oint_{c_0}\eta_t^*(-\mathbf{d}\nu+\operatorname{ad}_\nu\gamma)=\oint_{c_t}\operatorname{ad}_\nu\gamma\in\mathfrak{o}.
\]


\section{Applications}

In this section, we apply the processes of affine Euler-Poincar\'e and Lie-Poisson reductions to Yang-Mills magnetohydrodynamics and superfluids. The theory developed in this paper applies to a wide range of fluids with internal degrees of freedom, including microfluids, spin glasses, and liquid crystals. These examples are studied in detail in \cite{GBRa2008b}.

\subsection{Yang-Mills Magnetohydrodynamics}

Magnetohydrodynamics models the motion of an electrically charged and perfectly conducting fluid. In the balance of momentum law, one must add the Lorentz force of the magnetic field created by the fluid in motion. In addition, the hypothesis of infinite conductivity leads one to the conclusion that magnetic lines are frozen in the fluid, i.e. that they are transported along the particle paths. This hypothesis leads to the equation
\[
\frac{\partial}{\partial t}B+\boldsymbol{\pounds}_\mathbf{u}B=0.
\]
This model can be extend to incorporate nonabelian Yang-Mills interactions and is known under the name of Yang-Mills magnetohydrodynamics; see \cite{HoKu1984b} for a derivation of this model. Recall that for Yang-Mills theory, the field $B$ is seen as the curvature of a connection $A$ on a principal bundle. Clearly the connection $A$ represents the variable $\gamma$ in the general theory developed previously, on which the automorphism group acts by affine transformations. This shows that the abstract formalism developed previously is very natural in the context of Yang-Mills theory. Note that there is a more general model of fluid motion with Yang-Mills charged particles, namely the Euler-Yang-Mills equations. The Hamiltonian structure of these equations is given in \cite{GiHoKu1983}, see also \cite{GBRa2008a} for the associated Lagrangian and Hamiltonian reductions.

As remarked in \cite{HoKu1988}, at the reduced level, the Hamiltonian structure of Yang-Mills magnetohydrodynamics is given by the matrix \eqref{hamiltonian_matrix}. In this paragraph we carry out the corresponding affine Lie-Poisson reduction.

The group $G$ is chosen to be the semidirect product of the diffeomorphism group with the group of $\mathcal{O}$-valued function on $\mathcal{D}$, that is, $G=\operatorname{Diff}(\mathcal{D})\,\circledS\,\mathcal{F}(\mathcal{D},\mathcal{O})$. The order parameter Lie group $\mathcal{O} $ represents here the \textit{symmetry group of the particles interaction}. For example, $\mathcal{O}=S^1$ corresponds to electromagnetism, $
\mathcal{O}=SU(2)$ and $\mathcal{O}=SU(3)$ correspond to weak and strong interactions, respectively. The advected quantities are the \textit{mass density} $\rho$, the \textit{entropy density} $S$, and the \textit{potential of the Yang-Mills field} $A$. Therefore, we set
\[
a=(\rho,S)\in V_1^*=\mathcal{F}(\mathcal{D})\times\mathcal{F}(\mathcal{D}) \quad\text{and}\quad A=\gamma\in\Omega^1(\mathcal{D},\mathfrak{o}).
\]
The action of $(\eta,\chi)\in G$ on $(\rho,S)\in V_1^*$ is the usual right representation of the fluid relabeling group on the mass density and entropy density. It is given by
\[
(\rho,S)(\eta,\chi)=J\eta(\rho\circ\eta,S\circ\eta).
\]
The right affine action of $(\eta,\chi)\in G$ on $A\in \Omega^1(M,\mathfrak{o})$ is given, as in the example \eqref{affine_representation_gamma}, by
\[
A \mapsto  
\operatorname{Ad}_{\chi^{-1}}\eta^*A+\chi^{-1}T\chi.
\]
Since the variable $\kappa\in\mathcal{F}(\mathcal{D},\mathfrak{o}^*)$ is interpreted as the \textit{gauge-charge density}, we use the notations $Q\in\mathcal{F}(\mathcal{D},\mathfrak{o}^*)=T_0^*\mathcal{F}(\mathcal{D},\mathcal{O})$ and $Q_\chi\in T_\chi^*\mathcal{F}(\mathcal{D},\mathcal{O})$.

The reduced Hamiltonian $h:\Omega^1(\mathcal{D})\times\mathcal{F}(\mathcal{D},\mathfrak{o}^*)\times V_1^*\times V_2^*\rightarrow\mathbb{R}$ is given by
\[
h(\mathbf{m},Q,\rho,S,A)=\frac{1}{2}\int_\mathcal{D}\frac{1}{\rho} \|\mathbf{m}\|^2\mu+\int_\mathcal{D} \varepsilon(\rho,S)\mu+\frac{1}{2}\int_\mathcal{D}\|\mathbf{d}^AA\|^2\mu,
\]
where $\varepsilon$ denotes the \textit{internal energy density}, the norm in the first term is associated to a Riemannian metric $g$ on $\mathcal{D}$, and the norm in the third term is associated to the metric $(gk)$, on the vector bundle of $\mathfrak{o}$-valued $k$-forms on $\mathcal{D}$, induced by the metric $g$ and by an $\operatorname{Ad}$-invariant inner product $k$ on $\mathfrak{o}$. The metric $(gk)$ can be used to identify $\Omega_k(\mathcal{D},\mathfrak{o}^*)$ and its dual $\Omega^k(\mathcal{D},\mathfrak{o})$, by raising and lowering indices. Through this identification, the operators $\operatorname{div}$ and $\operatorname{div}^A$ act also on $\Omega^k(\mathcal{D},\mathfrak{o})$.

The affine Lie-Poisson equations \eqref{ALP_PerfectComplexFluid} associated to this Hamiltonian are
\begin{equation}\label{ALP_YMMHD_A}
\left\lbrace
\begin{array}{ll}
\vspace{0.2cm}\displaystyle\frac{\partial}{\partial t} \mathbf{u}+\nabla_\mathbf{u}\mathbf{u}=-\frac{1}{\rho}\left(\operatorname{grad}p+(gk)\left(\mathbf{i}_{\_\,}B,\operatorname{div}^AB\right)^\sharp\right),\quad B=\mathbf{d}^AA,\\
\vspace{0.2cm}\displaystyle\frac{\partial}{\partial t}Q+\operatorname{div}(Q\mathbf{u})=0,\qquad\displaystyle\frac{\partial}{\partial t}A+\mathbf{d}^A(A(\mathbf{u}))+\mathbf{i}_\mathbf{u}B=0,\\
\vspace{0.2cm}\displaystyle\frac{\partial}{\partial t}\rho+\operatorname{div}(\rho \mathbf{u})=0,\qquad\,\,\,\displaystyle\frac{\partial}{\partial t}S+\operatorname{div}(S\mathbf{u})=0,
\end{array} \right.
\end{equation}
where $\mathbf{m}=\rho\mathbf{u}^\flat$ and $p=\rho\mu_{\rm chem}+ST-\varepsilon$ is the \textit{pressure}, given in terms of the \textit{chemical potential} $\mu_{\rm chem}=\partial \varepsilon /\partial \rho$ and the \textit{temperature} $T=\partial \varepsilon/\partial S$. The first equation admits the stress tensor formulation
\[
\dot{\mathbf{m}}=-\operatorname{Div}\mathbf{T},
\]
where $\mathbf{T}$ is the $(1,1)$ stress tensor given by
\[
\mathbf{T}=\mathbf{u}\otimes\rho\mathbf{u}^\flat+B\!\cdot\!B+q\delta,\quad q=p-\frac{1}{2}\|B\|^2.
\]

We now treat the particular case of magnetohydrodynamics, that is, the case $\mathcal{O}=S^1$. In order to recover the standard equations we suppose that $\mathcal{D}$ is three dimensional. In this case we can define the \textit{magnetic potential} $\mathbf{A}:=A^\sharp \in \mathfrak{X}( \mathcal{D})$ and the \textit{magnetic field} $\mathbf{B}:=(\star B)^\sharp \in \mathfrak{X}( \mathcal{D}) $. Since the group is Abelian, covariant differentiation coincides with usual differentiation and the equality $\mathbf{d}^AA=B$ reads $\operatorname{curl}\mathbf{A}=\mathbf{B}$. Using the identities $(\operatorname{div}B)^\sharp=-\operatorname{curl}\mathbf{B}$ and $(\mathbf{i}_\mathbf{u}B)^\sharp=\mathbf{B}\times\mathbf{u}$ we obtain
\[
g\left(\mathbf{i}_{\_\,}B,\operatorname{div}B\right)^\sharp=-\left(\mathbf{i}_{(\operatorname{div}B)^\sharp}B\right)^\sharp=\mathbf{B}\times\operatorname{curl}\mathbf{B}.
\]
Suppose that all particles have mass $m$. The electric charge $q$ is such that $Q=\rho\frac{q}{m}$, therefore the equation for $Q$ in \eqref{ALP_YMMHD_A} becomes
\[
\frac{\partial}{\partial t}q+\mathbf{d}q(\mathbf{u})=0.
\]
If we suppose that at time $t=0$ all the particle have the same charge, then this charge remains constant for all time. Using these remarks and hypotheses, equations \eqref{ALP_YMMHD_A} become
\begin{equation}\label{MHD_A}
\left\lbrace
\begin{array}{ll}
\vspace{0.2cm}\displaystyle\frac{\partial}{\partial t} \mathbf{u}+\nabla_\mathbf{u}\mathbf{u}=-\frac{1}{\rho}\left(\operatorname{grad}p+\mathbf{B}\times\operatorname{curl}\mathbf{B}\right),\quad \mathbf{B}=\operatorname{curl}\mathbf{A},\\
\vspace{0.2cm}\displaystyle\frac{\partial}{\partial t}\mathbf{A}+\operatorname{grad}[g(\mathbf{A},\mathbf{u})]+\mathbf{B}\times\mathbf{u}=0,\\
\vspace{0.2cm}\displaystyle\frac{\partial}{\partial t}\rho+\operatorname{div}(\rho \mathbf{u})=0,\qquad\,\,\,\displaystyle\frac{\partial}{\partial t}S+\operatorname{div}(S\mathbf{u})=0.
\end{array} \right.
\end{equation}
Thus, we have recovered the equations for magnetohydrodynamics.

Turning back to the general case and using Theorem \ref{ALPSD}, we obtain the following result.

\paragraph{Hamiltonian reduction for Yang-Mills magnetohydrodynamics.} Consider the unreduced right invariant Hamiltonian $H(\mathbf{m}_\eta,\rho_\chi,S,\mathbf{v}_s)=H_{(S,\mathbf{v}_s)}(\mathbf{m}_\eta,\rho_\chi)$,
\[
H_{(S,\mathbf{v}_s)}:T^*(\operatorname{Diff}(\mathcal{D})\,\circledS\,\mathcal{F}(\mathcal{D},S^1))\rightarrow\mathbb{R},
\]
whose value at the identity is given by $h$. A smooth path $(\mathbf{m}_\eta,Q_\chi)\in T^*[\operatorname{Diff}(\mathcal{D})\,\circledS\,\mathcal{F}(\mathcal{D},\mathcal{O})]$ is a solution of Hamilton's equations associated to the Hamiltonian $H_{(\rho_0,S_0,A_0)}$ if and only if the curve
\[
(\rho\mathbf{u}^\flat,Q)=:(\mathbf{m},Q):=J(\eta^{-1})\left(\mathbf{m}_\eta\circ\eta^{-1},T^*R_{\chi\circ\eta^{-1}}(Q_\chi\circ\eta^{-1})\right)
\]
is a solution of the system \eqref{ALP_YMMHD_A}
with initial conditions $(\rho_0,S_0,A_0)$.

The evolution of the advected quantities is given by
\begin{align*}
\rho=J(\eta^{-1})(\rho_0\circ\eta^{-1}),\;\;S=J(\eta^{-1})(S_0\circ\eta^{-1}),\;\;A&=\operatorname{Ad}_{\chi\circ\eta^{-1}}\eta_*A_0+(\chi\circ\eta^{-1})T(\chi^{-1}\circ\eta^{-1})\\
&=\eta_*\left(\operatorname{Ad}_\chi A_0+\chi T\chi^{-1}\right).
\end{align*}
This theorem is interesting from two points of view. Firstly, it allows us to recover the non-canonical Hamiltonian structure given in \cite{HoKu1988} by a reduction from a canonical cotangent bundle. Secondly, it generalizes to the nonabelian case the Hamiltonian reduction for magnetohydrodynamics given in \cite{MaRaWe1984}.

The associated affine Lie-Poisson bracket is that given in \eqref{ALPB}, where the third term takes the explicit form
\[
\int_\mathcal{D}\rho\left(\textbf{d}\left(\frac{\delta f}{\delta
\rho}\right)\frac{\delta g}{\delta 
\mathbf{m}}-\textbf{d}\left(\frac{\delta g}{\delta
\rho}\right)\frac{\delta f}{\delta  \mathbf{m}}\right)\mu+\int_\mathcal{D}S\left(\textbf{d}\left(\frac{\delta f}{\delta
S}\right)\frac{\delta g}{\delta 
\mathbf{m}}-\textbf{d}\left(\frac{\delta g}{\delta
S}\right)\frac{\delta f}{\delta  \mathbf{m}}\right)\mu.
\]
In the general case of Yang-Mills magnetohydrodynamics, the Kelvin-Noether theorem gives
\[
\frac{d}{dt}\oint_{c_t}\mathbf{u}^\flat=\oint_{c_t}T\mathbf{d}s-\oint_{c_t}\frac{1}{\rho}(gk)\left(\mathbf{i}_{\_,}B,\operatorname{div}^AB\right),
\]
where $s$ is the \textit{specific entropy}. The $\gamma$-circulation gives
\[
\frac{d}{dt}\oint_{c_t}A=0,
\]
where $c_t$ is a loop which move with the fluid velocity $\mathbf{u}$, that is, $c_t=\eta_t \circ  c_0$.


\subsection{Hall Magnetohydrodynamics}

As we will see, Hall magnetohydrodynamics does not require the use of the affine Lie-Poisson reduction developed in this paper. However, in view of the next paragraph about superfluids, we quickly recall here from \cite{Ho1987} the Hamiltonian formulation of these equations. We will obtain this Hamiltonian structure by a Lie-Poisson reduction for semidirect products, associated to the direct product group $G:=\operatorname{Diff}(\mathcal{D})\times\operatorname{Diff}(\mathcal{D})$. The advected quantities are
\[
(\rho,S;n)\in\mathcal{F}(\mathcal{D})\times\mathcal{F}(\mathcal{D})\times\mathcal{F}(\mathcal{D}).
\]
The variables $\rho$ and $S$ are, as before, the \textit{mass density} and the \textit{entropy density}, on which only the first diffeomorphism group acts as
\[
(\rho,S)\mapsto(J\eta)(\rho\circ\eta,S\circ\eta).
\]
The variable $n$ is the \textit{electron charge density}, on which only the second diffeomorphism group acts as
\[
n\mapsto (J\xi)(n\circ\xi).
\]
By Lie-Poisson reduction, for a Hamiltonian $h=h(\mathbf{m},\rho,S;\mathbf{n},n)$ defined on the dual Lie-algebra
\[
\big([\mathfrak{X}(\mathcal{D})\,\circledS\,(\mathcal{F}(\mathcal{D})\times\mathcal{F}(\mathcal{D}))]\times[\mathfrak{X}(\mathcal{D})\,\circledS\,\mathcal{F}(\mathcal{D})]\big)^*\cong[\Omega^1(\mathcal{D})\times\mathcal{F}(\mathcal{D})\times\mathcal{F}(\mathcal{D})]\times[\Omega^1(\mathcal{D})\times\mathcal{F}(\mathcal{D})],
\]
we obtain the coupled Lie-Poisson equations
\begin{equation}\label{LP_H-MHD_1}
\left\{
\begin{array}{ll}
\vspace{0.1cm}\displaystyle\frac{\partial}{\partial t}\mathbf{m}=-{\boldsymbol{\pounds}}_{\frac{\delta h}{\delta \mathbf{m}}}\mathbf{m}-\operatorname{div}\left(\frac{\delta h}{\delta \mathbf{m}}\right)\mathbf{m}-\frac{\delta h}{\delta \rho}\diamond \rho-\frac{\delta h}{\delta S}\diamond S\\
\vspace{0.1cm}\displaystyle\frac{\partial}{\partial t}\rho=-\operatorname{div}\left(\frac{\delta h}{\delta \mathbf{m}}\rho\right)\\
\vspace{0.1cm}\displaystyle\frac{\partial}{\partial t}S=-\operatorname{div}\left(\frac{\delta h}{\delta \mathbf{m}}S\right)
\end{array}
\right.
\end{equation}
and
\begin{equation}\label{LP_H-MHD_2}
\left\{
\begin{array}{ll}
\vspace{0.1cm}\displaystyle\frac{\partial}{\partial t}\mathbf{n}=-{\boldsymbol{\pounds}}_{\frac{\delta h}{\delta \mathbf{n}}}\mathbf{n}-\operatorname{div}\left(\frac{\delta h}{\delta \mathbf{n}}\right)\mathbf{n}-\frac{\delta h}{\delta n}\diamond n\\
\vspace{0.1cm}\displaystyle\frac{\partial}{\partial t}n=-\operatorname{div}\left(\frac{\delta h}{\delta \mathbf{n}}n\right).
\end{array}
\right.
\end{equation}
The Hamiltonian for Hall magnetohydrodynamics is
\begin{equation}\label{Hamiltonian_H-MHD}
h(\mathbf{m},\rho,S;\mathbf{n},n):=\frac{1}{2}\int_\mathcal{D}\frac{1}{\rho}\left\|\mathbf{m}-\frac{a\rho}{R}A\right\|^2\mu+\int_\mathcal{D}\varepsilon(\rho,S)\mu+\frac{1}{2}\int_\mathcal{D}\|\mathbf{d}A\|^2\mu,
\end{equation}
where the one-from $A$, defined by
\[
A:=R\frac{\mathbf{n}}{n} \in\Omega^1(\mathcal{D}),
\]
is the \textit{magnetic vector potential}, the constants $a, R$ are respectively the \textit{ion charge-to-mass ratio} and the \textit{Hall scaling parameter}, and the norms are taken with respect to a fixed Riemannian metric $g$ on $\mathcal{D}$. Using the advection equations for $\rho$ and $n$ we obtain
\[
\frac{\partial}{\partial t}(a\rho+n)=0.
\]
Thus, if we assume that $a\rho_0+n_0=0$ for the initial conditions, we have $a\rho+n=0$ for all time. Using the Hamiltonian \eqref{Hamiltonian_H-MHD}, equations \eqref{LP_H-MHD_1} and \eqref{LP_H-MHD_2} are computed to be
\begin{equation}\label{H-MHD}
\left\{
\begin{array}{ll}
\vspace{0.1cm}\displaystyle\frac{\partial}{\partial t}\mathbf{u}+\nabla_\mathbf{u}\mathbf{u}=-\frac{1}{\rho}\left(\operatorname{grad}p-\left(\mathbf{i}_{(\operatorname{div}B)^\sharp}B\right)^\sharp\right)\\
\vspace{0.1cm}\displaystyle\frac{\partial}{\partial t}\rho=-\operatorname{div}\left(\rho\mathbf{u}\right),\quad\displaystyle\frac{\partial}{\partial t}S=-\operatorname{div}\left(S\mathbf{u}\right)\\
\vspace{0.1cm}\displaystyle\frac{\partial}{\partial t}A=-\mathbf{i}_\mathbf{u}B-\frac{R}{a\rho}\mathbf{i}_{(\operatorname{div}B)^\sharp}B,
\end{array}
\right.
\end{equation}
where $\mathbf{m}=\rho\mathbf{u}^\flat+\frac{a\rho}{R}A$ and $p=\rho\mu_{\rm chem}+ST-\varepsilon$ is the \textit{pressure}. The first equation admits the stress tensor formulation
\[
\dot{\mathbf{m}}+\dot{\mathbf{n}}=-\operatorname{Div}\mathbf{T},
\]
where $\mathbf{T}$ is the $(1,1)$ stress tensor given by
\[
\mathbf{T}=\mathbf{u}\otimes\rho\mathbf{u}^\flat+B\!\cdot\!B+q\delta,\quad q=p-\frac{1}{2}\|B\|^2.
\]
When $\mathcal{D}$ is three dimensional, we can define the \textit{magnetic potential} $\mathbf{A}:=A^\sharp$ and the \textit{magnetic field} $\mathbf{B}:=(\star B)^\sharp$. In this case the previous equations read
\begin{equation}\label{H-MHD_3D}
\left\{
\begin{array}{ll}
\vspace{0.1cm}\displaystyle\frac{\partial}{\partial t}\mathbf{u}+\nabla_\mathbf{u}\mathbf{u}=-\frac{1}{\rho}\left(\operatorname{grad}p+\mathbf{B}\times\operatorname{curl}\mathbf{B}\right)\\
\vspace{0.1cm}\displaystyle\frac{\partial}{\partial t}\rho=-\operatorname{div}\left(\rho\mathbf{u}\right),\quad\displaystyle\frac{\partial}{\partial t}S=-\operatorname{div}\left(S\mathbf{u}\right)\\
\vspace{0.1cm}\displaystyle\frac{\partial}{\partial t}\mathbf{A}=\mathbf{u}\times \mathbf{B}+\frac{R}{a\rho}\mathbf{B}\times\operatorname{curl}\mathbf{B}.
\end{array}
\right.
\end{equation}
These are the classical equations of Hall magnetohydrodynamics. Note that we can pass from the equations for magnetohydrodynamics to those for Hall magnetohydrodynamics by simply replacing the advection law for $A$ by the \textit{Ohm's law}. In terms of the magnetic field $B$, one simply replace the advection law
\[
\frac{\partial}{\partial t}B+\boldsymbol{\pounds}_\mathbf{u}B=0,
\]
where $\mathbf{u}$ is the fluid velocity, by the equation
\[
\frac{\partial}{\partial t}B+\boldsymbol{\pounds}_\mathbf{v}B=0,
\]
where $\mathbf{v}=\mathbf{u}+\frac{R}{a\rho}(\operatorname{div}B)^\sharp$ is the \textit{electron fluid velocity}.

\paragraph{Hamiltonian reduction for Hall magnetohydrodynamics.} Consider the unreduced right invariant Hamiltonian $H(\mathbf{m}_\eta,\rho,S;\mathbf{n}_\xi,n)=H_{(\rho,S;n)}(\mathbf{m}_\eta,\mathbf{n}_\xi)$,
\[
H_{(\rho,S;n)}:T^*(\operatorname{Diff}(\mathcal{D})\times\operatorname{Diff}(\mathcal{D}))\rightarrow\mathbb{R},
\]
whose value at the identity is given by $h$. Suppose that $a\rho_0+n_0=0$. A smooth path $(\mathbf{m}_\eta,\mathbf{n}_\xi)\in T^*(\operatorname{Diff}(\mathcal{D})\times\operatorname{Diff}(\mathcal{D}))$ is a solution of Hamilton's equations associated to $H_{(\rho_0,S_0;n_0)}$ if and only if the curve
\[
(\mathbf{m},\mathbf{n}):=(J(\eta^{-1})(\mathbf{m}_\eta\circ\eta^{-1}),J(\xi^{-1})(\mathbf{n}_\xi\circ\xi^{-1}))
\]
is a solution of the equations \eqref{H-MHD} where $A:=\frac{R}{n}\mathbf{n}=-\frac{R}{a\rho}\mathbf{n}$, since $a\rho+n=0$. Moreover the evolution of the advected quantities is given by
\[
\rho=J(\eta^{-1})(\rho_0\circ\eta^{-1}),\quad S=J(\eta^{-1})(S_0\circ\eta^{-1}),\quad n=J(\xi^{-1})(n_0\circ\xi^{-1}).
\]
Let us assume from now on that the initial conditions $\rho_0$ and  $n_0$ are related by $a \rho_0 + n _0 = 0 $. We have seen that this implies that  $a\rho+n=0$. From the relations above we conclude the interesting result that the action of $\xi^{-1}\circ\eta$ fixes $n_0$, that is, $J(\xi^{-1}\circ \eta)(n_0 \circ \xi^{-1}\circ \eta) = n_0$. Conversely, given this relation and the condition $a \rho_0 + n _0 = 0 $, it is
easily seen that $a\rho
+n=0$. 

The Lie-Poisson bracket associated to these equations is clearly the sum of two Lie-Poisson brackets associated to the semidirect products $\operatorname{Diff}(\mathcal{D})\,\circledS\,[\mathcal{F}(\mathcal{D})\times\mathcal{F}(\mathcal{D})]$ and $\operatorname{Diff}(\mathcal{D})\,\circledS\,\mathcal{F}(\mathcal{D})$.

The Kelvin-Noether theorem associated to the variable $\mathbf{m}$ gives
\[
\frac{d}{dt}\oint_{c_t}\left(\mathbf{u}^\flat+\frac{a}{R}A\right)=\oint_{c_t}T\mathbf{d}s,
\]
which can be rewritten as
\[
\frac{d}{dt}\oint_{c_t}\mathbf{u}^\flat=\oint_{c_t}T\mathbf{d}s+\oint_{c_t}\frac{1}{\rho}\mathbf{i}_{(\operatorname{div}B)^\sharp}B,
\]
where $c_t$ is a loop which moves with the \textit{fluid velocity} $\mathbf{u}$, that is, $c_t=\eta_t \circ c_0$, and $s$ is the \textit{specific entropy}. The Kelvin-Noether theorem associated to the variable $\mathbf{n}$ gives
\[
\frac{d}{dt}\oint_{d_t}A=0,
\]
where $d_t$ is a loop which moves with the electron fluid velocity $\mathbf{v}$, that is, $d_t=\xi_t \circ d_0$.


\subsection{Superfluids}
\label{sec:multivelocity}

Superfluidity is a rare state of matter encountered in few fluids at extremely low temperatures. Such materials exhibit strange behavior such as the lack of viscosity, the ability to flow through very small channels that are impermeable to ordinary fluids, and the fact that it can form a layer whose thickness is that of one atom on the walls of the container in which it is placed. In addition, the rotational speed of a superfluid is quantized, that is, the fluid can rotate only at certain values of the speed. Superfluidity is considered to be a manifestation of quantum mechanical effects at macroscopic level. Typical examples of superfluids are ${}^3$He, whose atoms are fermions and the superfluid transition occurs by Cooper pairing between atoms rather than electrons, and ${}^4$He, whose atoms are bosons and the superfluidity is a consequence of Bose-Einstein condensation in an interacting system.

For example at temperatures close to absolute zero a solution of ${}^3$He and ${}^4$He has its hydrodynamics described by three velocities: two superfluid velocities $\mathbf{v}_s ^1, \mathbf{v}_s^2 $ and one normal fluid velocity $\mathbf{v}_n $. If other kinds of superfluid are present, one needs to introduce additional superfluid velocities. For a history of the equations considered below and the Hamiltonian structure for superfluids see \cite{HoKu1987}.

For simplicity we treat the \textit{two-fluid model}, that is, the case of one superfluid velocity $\mathbf{v}_s$ and one normal-fluid velocity $\mathbf{v}_n$. Remarkably, this Hamiltonian structure can be obtained by affine Lie-Poisson reduction, with order parameter Lie group $\mathcal{O}=S^1$.

The linear advected quantity is the \textit{entropy density} $S$ on which a diffeomorphism $\eta$ acts as
\[
S\mapsto (J\eta)(S\circ\eta).
\]
The affine advected quantity is the \textit{superfluid velocity} $\mathbf{v}_s$, on which the element $(\eta,\chi)\in\operatorname{Diff}(\mathcal{D})\,\circledS\,\mathcal{F}(\mathcal{D},S^1)$ acts as
\[
\mathbf{v}_s\mapsto \left(\eta^*\mathbf{v}_s^\flat+\mathbf{d}\chi\right)^\sharp.
\]
This action is simply the affine representation \eqref{affine_representation_gamma} for the Lie group $\mathcal{O}=S^1$. Here the advected quantity $\mathbf{v}_s $ is a vector field and not a one-form, since it represents a velocity and hence the formula  \eqref{affine_representation_gamma} was changed accordingly. As will be seen,  in this formalism, the \textit{mass density} does not appear as an advected quantity in the representation space $V^*$; it is a momentum, that is, one of the variables in the dual Lie algebra $\mathfrak{g}^\ast = \Omega^1(\mathcal{D}) \times \mathcal{F}( \mathcal{D})$.

The reduced Hamiltonian is
\begin{equation}\label{hamiltonian_superfluids} 
h(\mathbf{m},\rho,S,\mathbf{v}_s)=-\frac{1}{2}\int_\mathcal{D}\rho\|\mathbf{v}_n\|^2\mu+\int_\mathcal{D}(\mathbf{m}\cdot\mathbf{v}_n)\mu+\int_\mathcal{D}\varepsilon(\rho,S,\mathbf{v}_s-\mathbf{v}_n)\mu,
\end{equation}
where $\mathbf{v}_n=\mathbf{v}_n(\mathbf{m},\rho,S,\mathbf{v}_s)$ is the vector field defined by the implicit condition
\begin{equation}\label{definition_v_n}
\mathbf{m}-\rho\mathbf{v}_n^\flat=\frac{\partial\varepsilon}{\partial\mathbf{r}}(\rho,S,\mathbf{v}_s-\mathbf{v}_n).
\end{equation}
By the implicit function theorem, the above relation defines a unique vector field $\mathbf{v}_n$, provided the function $\varepsilon$ verifies the condition that
\[
u_x\mapsto\frac{\partial^2\varepsilon}{\partial\mathbf{r}^2}(r,s,v_x)\cdot u_x-ru_x
\]
is a bijective linear map. 

The vector field $\mathbf{v}_n$ is interpreted as the \textit{velocity of the normal flow}. The \textit{internal energy density} $\varepsilon$ is seen here as a function of three variables $\varepsilon=\varepsilon(\rho,S,\mathbf{r}):\mathbb{R}\times\mathbb{R}\times T\mathcal{D}\rightarrow\mathbb{R}$. We make the following definitions:
\begin{align*}
\mu_{\rm chem}&:=\frac{\partial\varepsilon}{\partial\rho}(\rho,S,\mathbf{v}_s-\mathbf{v}_n)\in\mathcal{F}(\mathcal{D}),\quad T:=\frac{\partial\varepsilon}{\partial S}(\rho,S,\mathbf{v}_s-\mathbf{v}_n)\in\mathcal{F}(\mathcal{D}),\\
\mathbf{p}&:=\frac{\partial\varepsilon}{\partial\mathbf{r}}(\rho,S,\mathbf{v}_s-\mathbf{v}_n)\in\Omega^1(\mathcal{D}).
\end{align*}
The interpretation of the quantities $\mu_{\rm chem}, T$, and $\mathbf{p}$ is obtained from the following thermodynamic derivative identity for the internal energy (superfluid first law):
\[
\mathbf{d}(\varepsilon(\rho,S,\mathbf{v}_s-\mathbf{v}_n))=\mu_{\rm chem}\mathbf{d}\rho+T\mathbf{d}S+\mathbf{p}\cdot\nabla_{\_\,}(\mathbf{v}_s-\mathbf{v}_n)\in\Omega^1(\mathcal{D}),
\]
where $\nabla$ denotes the Levi-Civita covariant derivative associated to the metric $g$. The function $\mu_{\rm chem}$ is the \textit{chemical potential}, $T$ is the \textit{temperature}, and $\mathbf{p}$ is the \textit{relative momentum density}.

The affine Lie-Poisson equations \eqref{ALP_PerfectComplexFluid} associated to the Hamiltonian \eqref{hamiltonian_superfluids} are computed to be
\begin{equation}\label{superfluids}
\left\{
\begin{array}{ll}
\vspace{0.1cm}\displaystyle\frac{\partial}{\partial t}\mathbf{m}=-\operatorname{Div}\mathbf{T},\\
\vspace{0.1cm}\displaystyle\frac{\partial}{\partial t}\rho+\operatorname{div}(\rho\mathbf{v}_n+\mathbf{p}^\sharp)=0,\qquad\frac{\partial}{\partial t}S+\operatorname{div}(S\mathbf{v}_n)=0,\\
\vspace{0.1cm}\displaystyle\frac{\partial}{\partial t}\mathbf{v}_s+\operatorname{grad}\left(g(\mathbf{v}_s,\mathbf{v}_n)+\mu_{\rm chem}-\frac{1}{2}\|\mathbf{v}_n\|^2\right)+\left(\mathbf{i}_{\mathbf{v}_n}\mathbf{d}\mathbf{v}_s^\flat\right)^\sharp=0,
\end{array}
\right.
\end{equation}
where $\mathbf{m}=\rho\mathbf{v}_n^\flat+\mathbf{p}$ and $\mathbf{T}$ is the superfluid stress tensor given by
\[
\mathbf{T}=\mathbf{v}_n\otimes\mathbf{m}+\mathbf{p}^\sharp\otimes\mathbf{v}_s^\flat+p\,\delta,\quad p:=-\varepsilon(\rho,S,\mathbf{v}_s-\mathbf{v}_n)+\mu_{\rm chem}\rho+ST.
\]
The last equation can be rewritten as
\begin{equation}\label{equ_v_s}
\frac{\partial}{\partial t}\mathbf{v}_s+\nabla_{\mathbf{v}_s}\mathbf{v}_s=-\operatorname{grad}\left(\mu_{\rm chem}-\frac{1}{2}\|\mathbf{v}_s-\mathbf{v}_n\|^2\right)+\left(\mathbf{i}_{\mathbf{v}_s-\mathbf{v}_n}\mathbf{d}\mathbf{v}_s^\flat\right)^\sharp
\end{equation}
When $\mathcal{D}$ is three dimensional, the last term reads
\[
\left(\mathbf{i}_{\mathbf{v}_s-\mathbf{v}_n}\mathbf{d}\mathbf{v}_s^\flat\right)^\sharp=\left(\star\mathbf{d}\mathbf{v}_s^\flat\right)^\sharp\times (\mathbf{v}_s-\mathbf{v}_n)=\operatorname{curl}\mathbf{v}_s\times(\mathbf{v}_s-\mathbf{v}_n)=(\mathbf{v}_n-\mathbf{v}_s)\times\operatorname{curl}\mathbf{v}_s.
\]
Thus we have recovered the equations (1a) - (1d) in \cite{HoKu1987}, in the particular case of the two-fluids model.

\paragraph{Hamiltonian reduction for superfluids.}  Consider the right-invariant Hamiltonian function $H(\mathbf{m}_\eta,\rho_\chi,S,\mathbf{v}_s)=H_{(S,\mathbf{v}_s)}(\mathbf{m}_\eta,\rho_\chi)$ induced by $h$. A curve
\[
(\mathbf{m}_\eta,\rho_\chi)\in T^*[\operatorname{Diff}(\mathcal{D})\,\circledS\,\mathcal{F}(\mathcal{D},S^1)]
\]
is a solution of Hamilton's equations associated to the superfluid Hamiltonian $H_{(S_0,\mathbf{v}_{s 0})}$ if and only if the curve
\[
(\mathbf{m},\rho):=J(\eta^{-1})\left(\mathbf{m}_\eta\circ\eta^{-1},\rho_\chi\circ\eta^{-1}\right)
\]
is a solution of the system \eqref{superfluids}
with initial conditions $(S_0,\mathbf{v}_{s 0})$.

The evolution of the advected quantities is given by
\[
S=J(\eta^{-1})(S_0\circ\eta^{-1})\quad\text{and}\quad\mathbf{v}_s=\left(\eta_*\left(\mathbf{v}_{s 0}^\flat+\mathbf{d}\chi^{-1}\right)\right)^\sharp.
\]
Note that the evolution of the \textit{superfluid vorticity} is given by
\[
\mathbf{d}\mathbf{v}_s^\flat=\eta_*\mathbf{d}\mathbf{v}_{s 0}^\flat,
\]
therefore, the irrotationality condition $\mathbf{d}\mathbf{v}_s=0$ ($\operatorname{curl}\mathbf{v}_s=0$ for the three dimensional case) is preserved.

The associated Poisson bracket for superfluids is
\begin{align}\label{superfluids_bracket}
\{f,g\}(\mathbf{m},\rho,S,\mathbf{v}_s)&=\int_\mathcal{D}\mathbf{m}\cdot\left[\frac{\delta f}{\delta\mathbf{m}},\frac{\delta g}{\delta\mathbf{m}}\right]\mu+\int_\mathcal{D}\rho\cdot\left(\mathbf{d}\frac{\delta f}{\delta\rho}\cdot\frac{\delta g}{\delta\mathbf{m}}-\mathbf{d}\frac{\delta g}{\delta\rho}\cdot\frac{\delta f}{\delta\mathbf{m}}\right)\mu\nonumber\\
&\quad+\int_\mathcal{D}S\cdot\left(\mathbf{d}\frac{\delta f}{\delta S}\cdot\frac{\delta g}{\delta\mathbf{m}}-\mathbf{d}\frac{\delta g}{\delta S}\cdot\frac{\delta f}{\delta\mathbf{m}}\right)\mu\\
&\quad+\int_\mathcal{D}\left[\left(\mathbf{d}\frac{\delta f}{\delta\rho}+{\boldsymbol{\pounds}}_{\frac{\delta f}{\delta\mathbf{m}}}\mathbf{v}_s^\flat\right)\cdot\frac{\delta g}{\delta\mathbf{v}_s}^\sharp-\left(\mathbf{d}\frac{\delta g}{\delta\rho}+{\boldsymbol{\pounds}}_{\frac{\delta g}{\delta\mathbf{m}}}\mathbf{v}_s^\flat\right)\cdot\frac{\delta f}{\delta\mathbf{v}_s}^\sharp\right]\mu\nonumber.
\end{align}

The $\gamma$-circulation gives
\[
\frac{d}{dt}\oint_{c_t}\mathbf{v}_s^\flat=0,
\]
where $c_t$ is a loop which moves with the \textit{normal fluid velocity} $\mathbf{v}_n$.


\subsection{Superfluid Yang-Mills Magnetohydrodynamics}

In this paragraph we combine the Hamiltonian structures of Yang-Mills magnetohydrodynamics and superfluid dynamics, to obtain a new physical model for the theory of superfluids Yang-Mills magnetohydrodynamics as well as the corresponding Hamiltonian structure. In the Abelian case we recover the theory and the Hamiltonian structure derived in \cite{HoKu1987}. We need a slight generalization of the geometric framework developed in \S\ref{Hamiltonian_PCF}, namely we consider the group semidirect product
\[
\operatorname{Diff}(\mathcal{D})\,\circledS\,(\mathcal{F}(\mathcal{D},\mathcal{O})\times\mathcal{F}(\mathcal{D},S^1)),
\]
where $\mathcal{F}(\mathcal{D},\mathcal{O})\times\mathcal{F}(\mathcal{D},S^1)$ is a direct product of groups on which $\operatorname{Diff}(\mathcal{D})$ acts as
\[
(\chi_1,\chi_2)\mapsto (\chi_1\circ\eta,\chi_2\circ\eta).
\]
The affine advected quantities are the \textit{potential of the Yang-Mills fluid} $A$ and the \textit{superfluid velocity} $\mathbf{v}_s$, on which $(\eta,\chi_1,\chi_2)$ acts as
\[
A\mapsto \operatorname{Ad}_{\chi_1^{-1}}\eta^*A+\chi_1^{-1}T\chi_1\quad\text{and}\quad \mathbf{v}_s\mapsto (\eta^*\mathbf{v}_s^\flat+\mathbf{d}\chi_2)^\sharp.
\]
The reduced Hamiltonian is defined on the dual of the Lie algebra
\[
[\mathfrak{X}(\mathcal{D})\,\circledS\,(\mathcal{F}(\mathcal{D},\mathfrak{o})\times\mathcal{F}(\mathcal{D}))]\,\circledS\,(\mathcal{F}(\mathcal{D})\times\Omega^1(\mathcal{D},\mathfrak{o})\times\mathfrak{X}(\mathcal{D}))
\]
and is given by
\[
h(\mathbf{m},Q,\rho,S,A,\mathbf{v}_s)=-\frac{1}{2}\int_\mathcal{D}\rho\|\mathbf{v}_n\|^2\mu+\int_\mathcal{D}(\mathbf{m}\cdot\mathbf{v}_n)\mu+\int_\mathcal{D}\varepsilon(\rho,S,\mathbf{v}_s-\mathbf{v}_n)\mu+\frac{1}{2}\int_\mathcal{D}\|\mathbf{d}^AA\|^2\mu,
\]
where the \textit{normal fluid velocity} $\mathbf{v}_n$ is defined as in \eqref{definition_v_n}. This is simply the Hamiltonian \eqref{hamiltonian_superfluids} plus the energy of the Yang-Mills field. The norms are respectively associated to the metrics $g$ and $(gk)$, where $g$ is a Riemannian metric on $\mathcal{D}$ and $k$ is an $\operatorname{Ad}$-invariant inner product on $\mathfrak{o}$. The affine Lie-Poisson  equations associated to this Hamiltonian are computed to be
\begin{equation}\label{YM_superfluids}
\left\{
\begin{array}{ll}
\vspace{0.1cm}\displaystyle\frac{\partial}{\partial t}\mathbf{m}=-\operatorname{Div}\mathbf{T},\qquad\frac{\partial}{\partial t}\rho+\operatorname{div}(\rho\mathbf{v}_n+\mathbf{p}^\sharp)=0,\\
\vspace{0.1cm}\displaystyle\frac{\partial}{\partial t}Q+\operatorname{div}(Q\mathbf{v}_n)=0,\qquad\frac{\partial}{\partial t}S+\operatorname{div}(S\mathbf{v}_n)=0,\\
\vspace{0.1cm}\displaystyle\frac{\partial}{\partial t}\mathbf{v}_s+\operatorname{grad}\left(g(\mathbf{v}_s,\mathbf{v}_n)+\mu_{\rm chem}-\frac{1}{2}\|\mathbf{v}_n\|^2\right)+\left(\mathbf{i}_{\mathbf{v}_n}\mathbf{d}\mathbf{v}_s^\flat\right)^\sharp=0,\\
\vspace{0.1cm}\displaystyle\frac{\partial}{\partial t}A+\mathbf{d}^A(A(\mathbf{v}_n))+\mathbf{i}_{\mathbf{v}_n}B=0,\quad B:=\mathbf{d}^AA,
\end{array}
\right.
\end{equation}
where $\mathbf{m}=\rho\mathbf{v}_n^\flat+\mathbf{p}$ and $\mathbf{T}$ is the $(1,1)$ stress tensor given by
\[
\mathbf{T}:=\mathbf{v}_n\otimes\mathbf{m}+\mathbf{p}^\sharp\otimes\mathbf{v}_s^\flat+B\!\cdot\! B+ q\,\delta,\quad q=-\varepsilon(\rho,S,\mathbf{v}_s-\mathbf{v}_n)+\mu_{\rm chem}\rho+ST-\frac{1}{2}\|B\|^2.
\]

The corresponding Hamiltonian reduction and affine Lie-Poisson bracket can be found as before and the evolutions of the advected quantities are given by
\[
S=J(\eta^{-1})(S_0\circ\eta^{-1}),\;\; A=\eta_*\left(\operatorname{Ad}_\chi A_0+\chi_1 T\chi_1^{-1}\right)\;\;\text{and}\;\;\mathbf{v}_s=\left(\eta_*\left(\mathbf{v}_{s 0}^\flat+\mathbf{d}\chi_2^{-1}\right)\right)^\sharp.
\]
The $\gamma$-circulation gives
\[
\frac{d}{dt}\oint_{c_t}\mathbf{v}_s^\flat=0, \quad \text{and}\quad \frac{d}{dt}\oint_{c_t}A=0,
\]
where $c_t$ is a loop which moves with the \textit{normal fluid velocity} $\mathbf{v}_n$.


\subsection{Superfluid Hall Magnetohydrodynamics}
\label{sec:HallMHD}

The Hamiltonian formulation of superfluid Hall magnetohydrodynamics is given in \cite{HoKu1987}. As one can guess, the Hamiltonian structure of these equations combines the Hamiltonian structures of Hall magnetohydrodynamics and of superfluids. This is still true at the group level and we will obtain the equations by affine Lie-Poisson reduction associated to the group
\[
G:=\left[\operatorname{Diff}(\mathcal{D})\,\circledS\,\mathcal{F}(\mathcal{D},S^1)\right]\times\operatorname{Diff}(\mathcal{D}).
\]
In this expression, the symbol $\times$ denotes the \textit{direct product} of the two groups. The advected quantities are
\[
(S,\mathbf{u};n)\in\mathcal{F}(\mathcal{D})\times\mathfrak{X}(\mathcal{D})\times\mathcal{F}(\mathcal{D}).
\]
The variable $S$ is the \textit{entropy density} of the normal flow, the other variables will be interpreted later. The action of $(\eta,\chi;\xi)\in G$ is given by
\[
(S,\mathbf{u};n)\mapsto(J\eta(S\circ\eta),(\eta^*\mathbf{u}^\flat+\mathbf{d}\chi)^\sharp;J\xi(n\circ\xi)).
\]
The resulting affine Lie-Poisson equations consist of two systems, the affine Lie-Poisson equations associated to the variables $(\mathbf{m},\rho,S,\mathbf{u})$ and the Lie-Poisson equations associated to the variables $(\mathbf{n},n)$.

The Hamiltonian of superfluid Hall magnetohydrodynamics is defined on the dual Lie algebra
\begin{align*}
&\big(\left[(\mathfrak{X}(\mathcal{D})\,\circledS\,\mathcal{F}(\mathcal{D}))\,\circledS\,(\mathcal{F}(\mathcal{D})\oplus\mathfrak{X}(\mathcal{D}))\right]\times[\mathfrak{X}(\mathcal{D})\,\circledS\,\mathcal{F}(\mathcal{D})] \big)^\ast \\
&\quad\cong \Omega^1(\mathcal{D})\times\mathcal{F}(\mathcal{D})\times\mathcal{F}(\mathcal{D})\times\Omega^1(\mathcal{D})\times\Omega^1(\mathcal{D})\times\mathcal{F}(\mathcal{D})
\end{align*}
and is given by
\begin{align}\label{Ham_SFHMFD}
h(\mathbf{m},\rho,S,\mathbf{u};\mathbf{n},n):=&-\frac{1}{2}\int_\mathcal{D}\rho\|\mathbf{v}_n\|^2\mu+\int_\mathcal{D}\left(\left(\mathbf{m}-\frac{a\rho}{R} A\right)\cdot\mathbf{v}_n\right)\mu\nonumber\\
&+\int_\mathcal{D}\varepsilon(\rho,S,\mathbf{v}_s-\mathbf{v}_n)\mu+\frac{1}{2}\int_\mathcal{D}\|\mathbf{d}A\|^2\mu,
\end{align}
where $\mathbf{v}_n$ is the \textit{velocity of the normal flow}, $\mathbf{v}_s:=\mathbf{u}-\frac{a}{R}A^\sharp$ is the \textit{superfluid velocity}, and $\varepsilon$ is the \textit{internal energy density}. The one-form $A$ is defined by
\[
A:=R\frac{\mathbf{n}}{n}.
\]
The norm in the first term is taken with respect to a fixed Riemannian metric $g$ on $\mathcal{D}$. The velocity $\mathbf{v}_n$ is the function $\mathbf{v}_n=\mathbf{v}_n(\mathbf{m},\rho,S,\mathbf{u};\mathbf{n},n)$ defined by the implicit condition
\[
\mathbf{m}-\rho\mathbf{v}_n^\flat-\frac{a\rho}{R}A=\frac{\partial\varepsilon}{\partial\mathbf{r}}(\rho,S,\mathbf{v}_s-\mathbf{v}_n).
\]
By the implicit function theorem, the above relation defines a unique function $\mathbf{v}_n$, provided the function $\varepsilon$ verifies the condition that the linear map
\[
u_x\mapsto\frac{\partial^2\varepsilon}{\partial\mathbf{r}^2}(r,s,v_x,w_x)\cdot u_x-ru_x
\]
is bijective for all $(r,s,v_x,w_x)\in\mathbb{R}\times\mathbb{R}\times T\mathcal{D}\times T\mathcal{D}$. Using the equations for $\rho$ and $n$ we obtain
\[
\frac{\partial}{\partial t}(a\rho+n)=0.
\]
Thus, if we assume that the initial conditions verify $a\rho_0+n_0=0$, then we have $a\rho+n=0$ for all time. In this case, the affine Lie-Poisson equations associated to the Hamiltonian \eqref{Ham_SFHMFD} are given by
\begin{equation}\label{superfluid_H_MHD}
\left\{
\begin{array}{ll}
\vspace{0.1cm}\displaystyle\frac{\partial}{\partial t}(\mathbf{m}+\mathbf{n})=-\operatorname{Div}\mathbf{T},\\
\vspace{0.1cm}\displaystyle\frac{\partial}{\partial t}\rho+\operatorname{div}(\rho\mathbf{v}_n+\mathbf{p}^\sharp)=0,\qquad\frac{\partial}{\partial t}S+\operatorname{div}(S\mathbf{v}_n)=0,\\
\vspace{0.1cm}\displaystyle\frac{\partial}{\partial t}A=-\mathbf{i}_{\mathbf{v}_n}B-\frac{1}{\rho}\mathbf{i}_{\mathbf{p}^\sharp}B-\frac{R}{a\rho}\mathbf{i}_{(\operatorname{div}B)^\sharp}B,\\
\vspace{0.1cm}\displaystyle\frac{\partial}{\partial t}\mathbf{v}_s=-\operatorname{grad}\left(g(\mathbf{v}_s,\mathbf{v}_n)+\mu_{\rm chem}-\frac{1}{2}\|\mathbf{v}_n\|^2\right)\\
\qquad \quad  \; \;+\left(\frac{a}{R\rho}\mathbf{i}_{\mathbf{p}^\sharp}B+\frac{1}{\rho}\mathbf{i}_{(\operatorname{div}B)^\sharp}B-\mathbf{i}_{\mathbf{v}_n}\mathbf{d}\mathbf{v}_s^\flat\right)^\sharp,
\end{array}
\right.
\end{equation}
where $\mathbf{m}=\rho\mathbf{v}_n^\flat+\frac{a\rho}{R}A+\mathbf{p}$ and $\mathbf{T}$ is $(1,1)$ stress tensor given by
\[
\mathbf{T}=\mathbf{v}_n\otimes (\rho\mathbf{v}_n^\flat+\mathbf{p})+\mathbf{p}^\sharp+B\!\cdot\!B+q\delta,\quad q=-\varepsilon(\rho,S,\mathbf{v}_s-\mathbf{v}_n)+\rho\mu_{\rm chem}+ST-\frac{1}{2}\|B\|^2.
\]
These are the equations for superfluid Hall magnetohydrodynamics as given in \cite{HoKu1987} equations (35a)--(35e). When $\mathcal{D}$ is three dimensional, the two last equations read
\[
\frac{\partial}{\partial t}A^\sharp=\left(\mathbf{v}_n+\frac{1}{\rho}\mathbf{p}^\sharp-\frac{R}{a\rho}\operatorname{curl}\mathbf{B}\right)\times\mathbf{B}\;\;\text{and}
\]
\[
\frac{\partial}{\partial t}\mathbf{v}_s=-\operatorname{grad}\left(g(\mathbf{v}_s,\mathbf{v}_n)+\mu_{\rm chem}-\frac{1}{2}\|\mathbf{v}_n\|^2\right)+\mathbf{v}_n\times\operatorname{curl}\mathbf{v}_s+\frac{1}{\rho}\left(\operatorname{curl}\mathbf{B}-\frac{a}{R}\mathbf{p}^\sharp\right)\times\mathbf{B}.
\]
\paragraph{Hamiltonian reduction for superfluid Hall magnetohydrodynamics.} Consider the right-invariant Hamiltonian function $H(\mathbf{m}_\eta,\rho_\chi,S,\mathbf{u};\mathbf{n}_\xi,n)=H_{(S,\mathbf{u};n)}(\mathbf{m}_\eta,\rho_\chi;\mathbf{n}_\xi)$ induced by $h$ and suppose that we have $a\rho_0+n_0=0$. A smooth curve
\[
(\mathbf{m}_\eta,\rho_\chi;\mathbf{n}_\xi)\in T^*\left[(\operatorname{Diff}(\mathcal{D})\,\circledS\,\mathcal{F}(\mathcal{D},S^1))\times\operatorname{Diff}(\mathcal{D})\right]
\]
is a solution of Hamilton's equations associated to $H_{(S_0,\mathbf{u}_0;n_0)}$ and with the initial condition $\rho_0$ if and only if the curve
\[
(\mathbf{m},\rho;\mathbf{n}):=\left(J(\eta^{-1})(\mathbf{m}_\eta\circ\eta^{-1}),J(\eta^{-1})(\rho_\chi\circ\eta^{-1});J(\xi^{-1})(\mathbf{n}\circ\xi^{-1})\right)
\]
is a solution of the equations \eqref{superfluid_H_MHD}, where $\mathbf{v}_s=\mathbf{u}- a A^\sharp/R=\mathbf{u}-a\mathbf{n}/n$.

The Poisson bracket for superfluid Hall magnetohydrodynamics is the sum of the affine Lie-Poisson bracket associated to the variables $(\mathbf{m},\rho,S,\mathbf{u})$ and the Lie-Poisson bracket associated to the variables $(\mathbf{n},n)$.

\medskip

The $\gamma$-circulation gives
\[
\frac{d}{dt}\oint_{c_t}\mathbf{u}^\flat=0.
\]
Using the definition $\mathbf{v}_s:=\mathbf{u}-\frac{a}{R}A^\sharp$, we obtain
\[
\frac{d}{dt}\oint_{c_t}\left(\mathbf{v}_s^\flat+\frac{a}{R}A^\sharp\right)=0,
\]
where $c_t$ is a loop which moves with the \textit{normal fluid velocity} $\mathbf{v}_n$. The Kelvin-Noether theorem associated to the variable $\mathbf{n}$ gives
\[
\frac{d}{dt}\oint_{d_t}A=0,
\]
where $d_t$ is a loop which moves with the \textit{electron fluid velocity} $\mathbf{v}$.



\paragraph{Acknowledgments.} The authors acknowledge the partial support of the Swiss National Science Foundation. Our special thanks go to Darryl Holm for his invaluable explanations of the physical phenomena described in the examples treated in this paper and for pointing out that a general abstract theory that would encompass all these examples was lacking. His patience and numerous clarifications were crucial in our understanding of these models.

{\footnotesize

\bibliographystyle{new}

\addcontentsline{toc}{section}{References}
\begin{thebibliography}{300}





\bibitem[Bia\l ynicki-Birula, Hubbard, and Turski(1983)]{BiHuTu1984}
Bia\l ynicki-Birula, I., J.~C. Hubbard, and  \L.~A. Turski [1984], Gauge-independent canonical formulation of relativistic plasma theory. \textit{Phys. A}, \textbf{128}(3), 509--519.



\bibitem[Cendra, Marsden, and Ratiu(2003)]{CeMaRa2003}
Cendra, H., J.~E. Marsden, and T.~S. Ratiu [2003], Cocycles, compatibility, and Poisson brackets for complex fluids, \textit{Advances in Multifield Theories of Continua with Substructure}, G. Capriz and P. M. Mariano (Eds.), Modeling and Simulation in Science, Engineering and Technology Series, Birkh\"auser, Boston, 51--73.





\bibitem[Gay-Balmaz and Ratiu(2008)]{GBRa2008a}
Gay-Balmaz, F. and T.~S. Ratiu [2008], Reduced Lagrangian and Hamiltonian formulations of Euler-Yang-Mills fluids, \textit{Journ. Sympl. Geom.}, \textbf{6} (2) 189--237. 


\bibitem[Gay-Balmaz and Ratiu(2008)]{GBRa2008b}
Gay-Balmaz, F. and T.~S. Ratiu [2008], The geometric structure of complex fluids,  \textit{Adv. Appl. Math.}, \textbf{42}(2), 176--275. 


\bibitem[Gibbons, Holm, and Kupershmidt(1983)]{GiHoKu1983}
Gibbons, J., D.~D. Holm and B.~A. Kupershmidt [1983], The Hamiltonian
Structure of Classical Chromohydrodynamics, \textit{Physica D},
\textbf{6}, 179--194.


\bibitem[Holm(1987)]{Ho1987}
Holm, D.~D. [1987], Hall magnetohydrodynamics : Conservation laws and Lyapunov stability, \textit{Phys. Fluids}, \textbf{30}, 1310--1322.

\bibitem[Holm(2001)]{Ho2001}
Holm, D.~D. [2001], Introduction to HVBK dynamics, \textit{Lecture Notes in Physics}, \textbf{571}, 114--130.

\bibitem[Holm(2002)]{Ho2002}
Holm, D.~D. [2002], Euler-Poincar\'e dynamics of perfect complex fluids, in \textit{Geometry, Dynamics and Mechanics: 60th Birthday Volume for J.E. Marsden.} P. Holmes, P. Newton, and A. Weinstein, eds., Springer-Verlag.

\bibitem[Holm and Kupershmidt(1984a)]{HoKu1984a}
Holm, D.~D. and B.~A. Kupershmidt [1984], Relativistic fluid dynamics as a Hamiltonian system. \textit{Phys. Lett. A}, \textbf{101}(1), 23--26.

\bibitem[Holm and Kupershmidt(1984b)]{HoKu1984b}
Holm, D.~D. and B.~A. Kupershmidt [1984], Yang-Mills magnetohydrodynamics: nonrelativistic theory, \textit{Phys. Rev. D}, \textbf{30}, 2557--2560.

\bibitem[Holm and Kupershmidt(1987)]{HoKu1987}
Holm, D.~D. and B.~A. Kupershmidt [1987], Superfluid plasmas: Multivelocity nonlinear hydrodynamics of superfluid solutions with charged condensates coupled electromagnetically, \textit{Phys. Rev. A}, \textbf{36}, 3947--3956.

\bibitem[Holm and Kupershmidt(1988)]{HoKu1988}
Holm, D.~D. and B.~A. Kupershmidt [1988], The analogy between spin glasses and Yang-Mills fluids, \textit{J. Math. Phys.} \textbf{29}, 21--30.

\bibitem[Holm, Marsden, and Ratiu(1998)]{HoMaRa1998}
Holm D.~D., J.~E. Marsden and T.~S. Ratiu [1998], The Euler-Poincar\'e equations and semidirect products with applications to continuum theories, \textit{Adv. in Math.}, \textbf{137}, 1--81.






\bibitem[Marsden et al(2007)]{MaMiOrPeRa2007}
Marsden, J.~E., G. Misio\l ek, J.-P. Ortega, M. Perlmutter, and T.~S. Ratiu [2007], \textit{Hamiltonian Reduction by Stages}, Springer Lecture Notes in Mathematics, \textbf{1913}, Springer-Verlag 2007.

\bibitem[Marsden and Ratiu(1999)]{MaRa1999} Marsden, J.~E.
and T.~S. Ratiu [1994], {\it Introduction to Mechanics and
Symmetry}, Texts in Applied Mathematics, {\bf 17},
Springer-Verlag, 1994;
\newblock Second Edition, 1999, second printing 2003.

\bibitem[Marsden, Ratiu, and Weinstein(1984)]{MaRaWe1984}
Marsden, J.~E., T.~S. Ratiu and A. Weinstein [1984], Semidirect product and
reduction in mechanics, \textit{Trans. Amer. Math. Soc.},
\textbf{281}, 147-177.

\bibitem[Marsden, Weinstein, Ratiu, Schmid, and Spencer(1983)]{MaWeRaScSp1983}
Marsden, J.~E., A. Weinstein, T.~S. Ratiu, R. Schmid and R.~G.
Spencer [1982], Hamiltonian system with symmetry, coadjoint orbits
and Plasma physics, in \textit{Proc. IUTAM-IS1MM Symposium on
Modern Developments in Analytical Mechanics (Torino 1982)},
\textbf{117}, 289-340, Atti della Acad. della Sc. di Torino.

\bibitem[Mayer(1984)]{Ma1984}
Mayer, M.~E. [1984], Poisson structures for relativistic systems (relativistic superfluids). In \textit{Fluids and plasmas: geometry and dynamics} (Boulder, Colo., 1983), 177--188, \textit{Contemp. Math.}, \textbf{28}, Amer. Math. Soc., Providence, RI.

\bibitem[Morrison and Greene (1980)]{MoGr1980}
Morrison, P.~J. and J.~M. Greene [1980], Noncanonical Hamiltonian density formulation of hydrodynamics and ideal magnetohydrodynamics, \textit{Phys. Rev. Lett.}, \textbf{45}, 790--794; errata \textbf{48} (1982), 569.











\end{thebibliography}

}

\end{document}